\documentclass[12pt,draftcls,onecolumn]{IEEEtran}\IEEEoverridecommandlockouts
\usepackage{etoolbox}
\makeatletter
\patchcmd{\@makecaption}
{\scshape}
{}
{}
{}
\makeatletter
\patchcmd{\@makecaption}
{\\}
{.\ }
{}
{}
\makeatother
\usepackage{amsmath,amssymb}
\usepackage{subcaption}
\usepackage{graphicx,graphics,color,psfrag}
\usepackage{cite,balance}
\usepackage{caption}
\allowdisplaybreaks
\usepackage{accents}
\usepackage{amsthm}
\usepackage{bm}
\usepackage[english]{babel}
\usepackage{multirow}
\usepackage{enumerate}
\usepackage{cases}
\usepackage{stfloats}
\usepackage{dsfont}
\usepackage{soul}
\usepackage{amsfonts}
\usepackage{fancyhdr}
\usepackage{hhline}
\usepackage{array}
\usepackage{float}
\usepackage{epstopdf}
\usepackage{calc}
\usepackage{diagbox}
\usepackage{verbatim}
\usepackage{booktabs}
\usepackage[ruled]{algorithm2e}

\begin{document}

\title{Age of Information-based Scheduling for Wireless D2D Systems with a Deep Learning Approach}
	\author{Ling Luo, Zhenyu Liu, Zhiyong Chen, Min Hua, Wenqing Li, and Bin Xia,~\IEEEmembership{Senior Member,~IEEE}
	\thanks{Ling Luo, Min Hua and Wenqing Li are with the State Grid Shanghai Electric Power Research Institute, Shanghai, China (e-mail: luol@sh.sgcc.com.cn; hmhzgb@163.com; xliwenqing\_2263@163.com).}
	\thanks{Zhenyu Liu, Zhiyong Chen and Bin Xia are with the Shanghai Jiao Tong University, Shanghai, China (e-mail: liuzhy@sjtu.edu.cn; zhiyongchen@sjtu.edu.cn; bxia@sjtu.edu.cn).}}

	\maketitle

\begin{abstract}
Device-to-device (D2D) links scheduling for avoiding excessive interference is critical to the success of wireless D2D communications. Most of the traditional scheduling schemes only consider the maximum throughput or fairness of the system and do not consider the freshness of information.
In this paper, we propose a novel D2D links scheduling scheme to optimize an age of information (AoI) and throughput jointly scheduling problem when D2D links transmit packets under the last-come-first-serve policy with packet-replacement(LCFS-PR).
It is motivated by the fact that the maximum throughput scheduling may reduce the activation probability of links with poor channel conditions, which results in terrible AoI performance.
Specifically, We derive the expression of the overall average AoI and throughput of the network under the spatio-temporal interfering queue dynamics with the mean-field assumption.
Moreover, a neural network structure is proposed to learn the mapping from the geographic location to the optimal scheduling parameters under a stationary randomized policy, where the scheduling decision can be made without estimating the channel state information(CSI) after the neural network is well-trained.
To overcome the problem that implicit loss functions cannot be back-propagated, we derive a numerical solution of the gradient.
Finally, numerical results reveal that the performance of the deep learning approach is close to that of a local optimal algorithm which has a higher computational complexity. The trade-off curve of AoI and throughput is also obtained, where the AoI tends to infinity when throughput is maximized.

\end{abstract}

\section{Introduction}
The booming Internet of Things (IoT) network has raised growing demand for real-time transmission of information, while the dense deployment of communication devices increasing rapidly creates a heavily interfering environment.
\textcolor{blue}{The Global System for Mobile Communications Association (GSMA)\cite{2021mobile} predicted that the number of connected IoT devices will increase from 13.1 billion in 2020 to 24 billion in 2025. Massive machine-type devices are connected to the cellular network, which increases the pressure on the network.
For massive machine-type communications, the existing frequency resources are not enough to realize orthogonal access for all devices.}
Device-to-device (D2D) communications, which is a key technology for supporting IoT scenarios, can reduce the pressure of the base station (BS) and enhance spectral efficiency through devices communicating directly in a frequency reuse pattern instead of via the BS.

Taking into account the scarcity of spectrum resources, most D2D communications operate in a frequency reuse model, yielding significant interference for each other when they are activated at the same time. It is obvious that D2D communication requires a careful schedule of D2D links. The main research on D2D scheduling of wireless D2D networks focuses on the analysis and optimization of sum-rate\cite{FPLinQ}, throughput\cite{CachingPolicy}, the user fairness\cite{Heuristic} or traffic density\cite{8688635}, but these metrics are insufficient to characterize the freshness of information.
The freshness of information is becoming more and more important in real-time services provided by wireless D2D communications, such as real-time monitoring. When a device needs to transmit information, but the D2D link is not scheduled, the freshness of the information decreases. Therefore, it is of great importance to design the D2D scheduling scheme in wireless D2D communications to get more fresh information.

Recently, the age of information (AoI) is proposed to characterize the freshness of information in \cite{Real-time}. AoI can measure the time elapsed since the latest received update was generated from the perspective of the receiver. \cite{Real-time} analyzes the AoI performance in some queuing systems such as $M/M/1$ by adjusting the arrival rate under a first-come-first-served (FCFS) policy. Following this, plenty of research has been conducted with regard to the analysis and minimization of the AoI in various systems, wherein methods include utilizing last-come-first-served (LCFS) policy and discarding obsolete packets\cite{OntheAge}, adjusting packet generation policy\cite{Update}, introducing packet deadlines to avoid prolonged blockage\cite{8323423} and preemption policy to ensure the package is up-to-date\cite{8445919} in the single source-destination system.
To extend the analysis of AoI to a more general and large-scale system, multiple sources at a single-server queue scenario is considered in\cite{8469047} and parallel-server queue such as $M/M/c$ under preemption policy are analyzed in\cite{8437907}. Considering the actual channel interference, AoI in the wireless communication system is also studied:\cite{8006541} analyzes the influence of code redundancy in erasure channel, carrier sense multiple access(CSMA) system under battery lifetime constraints are considered in\cite{10.1145/3397166.3409125}, online or offline scheduling policies to minimize the weighted sum average AoI are proposed in \cite{Scheduling} under the network only a single link can be scheduled to transmit in each slot.

Nevertheless, there are few works on the analysis and minimization of AoI in systems such as D2D where multiple direct communication links share the same spectrum and interfere with each other when they are activated simultaneously. Existing researches utilize stochastic geometry to model the spatial relationship of D2D devices\cite{mankar,OptimizingH,9042825}.
With this tool, \cite{mankar} statistically analyzes the distribution of the peak AoI and the conditional success probability in D2D scenarios, where two different queuing types are considered. \cite{OptimizingH} proposes a decentralized scheduling policy aiming at minimizing the peak AoI by receivers utilizing the channel state information(CSI) of the links around their stopping sets. \cite{9042825} considers the uplink network peak AoI under time-triggered and event-triggered traffic in an IoT scenario with mutual interference.

However, scheduling problems for the wireless networks with complex interference are usually non-convex and NP-hard\cite{7492912}, which drives researchers to employ machine learning and artificial intelligence to find a local optimal solution\cite{HORNIK1989359}. \cite{8664604} exploits a spatial convolution neural network to solve the sum-rate maximize problem in an all frequency reuse dense wireless network. \cite{9376717} compares several different deep reinforcement learning methods for AoI minimization under resource constraints of multi-user networks using hybrid automatic repeat requests. \cite{9097584} considers an IoT system where the application can be updated if all devices relevant to it are transmitted in a certain frame and using a deep Q network to learn the scheduling policy. \cite{wang2021distributed} propose a distributed reinforcement learning (RL) approach for the sampling policy to minimize the weighted sum of AoI cost and energy consumption, which allows the edge devices to cooperatively find the optimal policy with the local observations.

Motivated by this, we develop a novel D2D link scheduling policy based on deep learning for minimizing the overall average AoI of wireless D2D communications in this paper. Different from \cite{mankar} where the author set all links to the same activation probability and focus on the overall analysis of statistical spatial distribution, our scheduling policy can generate activation probability for each link to minimize AoI. \cite{OptimizingH} makes the scheduling decisions decentralized by scaling the problem excessively and unable to make decisions with other links jointly, while our method is absorbed in the optimization considering the global information of all links. Numerical results reveal that our deep learning algorithm performs better than these methods in AoI minimization task. Besides, our joint scheduling algorithm can assign different importances to AoI and throughput for differentiated demands. The main contributions of this paper are summarized as follows:

\begin{itemize}
	\item We analyze the transmission success probability in spatio-temporal correlated queue where the correlation is caused by spatial coupled interference and temporal dependent buffer state in the D2D network, in which path-loss and Rayleigh fading are taken into account. We derive the expression of the overall average AoI and throughput of the network under stationary randomized policy and formulate the AoI-throughput joint scheduling problem.
	
	\item Inspired by \cite{8664604}, we exploit a deep learning approach to solve the non-convex scheduling problem. A neural network structure is proposed to learn the mapping from specific geographic location information (GLI) to the corresponding scheduling decisions bypassing the CSI estimation. To enable the implicit loss function to be trained by the back-propagation, we derive a numerical solution to the gradients of this implicit expression.
	
	\item We utilize the proposed neural network algorithm to explore the trade-off between network throughput and AoI by taking the importance weight as a network input.
	After the neural network is well-trained, scheduling policy parameters can be output for disparate weight, the packet generating rate and D2D devices distribution.
	
	\item Numerical results show that the performance of the proposed AoI-based scheduling policy by using the neural network is close to that of the traditional algorithm under different network parameters, while the computational complexity is greatly reduced and CSI \textcolor{blue}{and path-loss are} not required. Besides, for a single link, AoI and throughput have no trade-off, while for a large-scale network, AoI tends to infinity when throughput is maximized. Scheduling to minimize AoI places greater emphasis on fairness, but sacrifices throughput performance. \textcolor{blue}{Finally, the optimal AoI is not achieved at the most frequent packet generation.}
\end{itemize}

The remainder of the paper is organized as follows: In Section \ref{system}, we establish the system model and introduce the stationary randomized policy. In Section \ref{problem}, we derive the expression of the D2D network state and formulate the joint scheduling optimization problem. In Section \ref{aoi}, we propose a deep learning method to solve this problem. Numerical results and conclusion are presented in Section \ref{numerical} and \ref{conclusion} respectively.


\section{System Model} \label{system}
\begin{figure}[t]
	\centering
	\includegraphics[width=10cm]{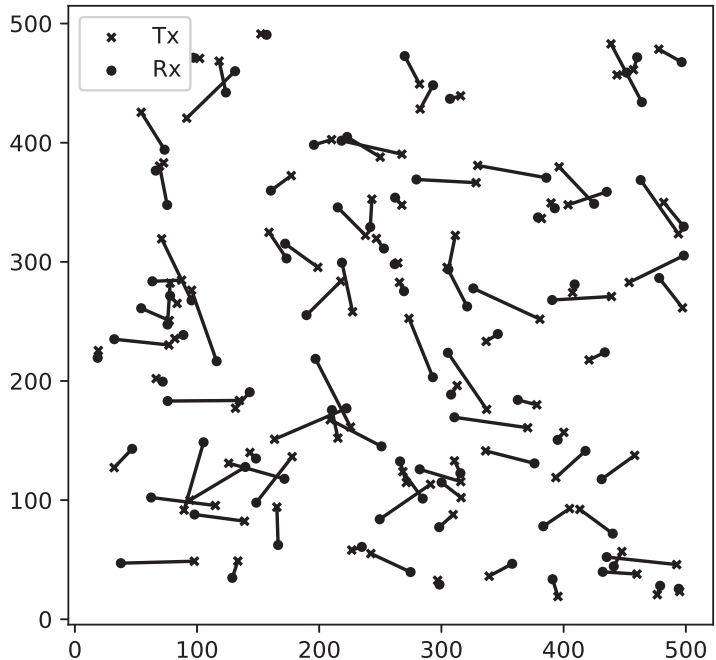}
	\caption{Multiple D2D pairs in wireless D2D communications.}
	\label{fig4}
\end{figure}
Consider a wireless network with $N$ independent D2D device links transmitting time-sensitive information, in which the devices are randomly distributed in a two-dimensional square region, and the distance between the transmitter and the receiver of each link varies, as shown in Fig. \ref{fig4}. We assume the network is static, that is, after the location initialization, the network topology remained unchanged. This assumption is reasonable since in many practical systems, devices are fixed after the deployment, and spatial dynamics are insignificant compared with the time scale.

The system operates in time-slotted fashion with time slot index $k\in\{1,2,\dots,K\} $. At the beginning of each time slot, new packets are generated in transmitters according to the Bernoulli process with the arrival rate $\xi \in (0,1]$. Each transmitter stores the up-to-date packet in a unit-size buffer and discard the undelivered one, which is called a last-come-first-serve policy with packet replacement (LCFS-PR) \cite{OntheAge}. Let $a_i(k)$ be an indicator function that is equal to 1 when link $i$ is scheduled to activate and $a_i(k)=0$ otherwise. Denote $\textbf{A}(k)=\{a_1(k), a_2(k),\dots, a_N(k) \}$ as the scheduling decisions of all $N$ links in time slot $k$. When scheduled to transmit, the transmitter will send a packet containing the real-time information after the packet generation process at the beginning of the slot. We assume that each time slot is long enough to transmit one packet and packets that fail to decode on the receiver will revert to the buffer. For the case where the buffer is empty, i.e., the buffered packet has been successfully transmitted and no fresher packet is generated, the link will not be activated. We use $n_i(k)$ to indicate the buffer state of link $i$ at slot $k$ with $n_i(k) = 1$ for non-empty and $n_i(k) = 0$ for empty. For empty buffer, each link is fixed inactive regardless of the value of $a_i(k)$.

We consider full frequency reuse, that is, all D2D links share the same frequency band \cite{8664604, yang2021spatiotemporal}.
\textcolor{blue}{This model is based on the fact that the number of orthogonal spectrum resource blocks is insufficient to allocate interference-free communication environments for all D2D links. In addition, due to the close distance of D2D communication devices, multiple links sharing the same resource block with appropriate interference management technology can improve spectral efficiency.}

Under the assumption of frequency reuse, when a subset of links is scheduled simultaneously, interference with each other reduces the successful probability of decoding at the receiver. Given a specific scheduling decision $\textbf{A}(k)$, the signal-to-interference-plus-noise ratio (SINR) measured at the receiver $i$ in the time slot $k$ can be written as\cite{OptimizingH}
\begin{equation}
	\mathrm{SINR}_i(k) = \frac {P_{tx}h_{ii}\lVert{d_{ii}}\rVert ^{-\alpha}a_i(k)n_i(k)}{\sum_{j\neq{i}}{P_{tx}h_{ji}\lVert{d_{ji}}\rVert ^{-\alpha}a_j(k)n_j(k)}+\sigma^2} , \label{0}
\end{equation}
where $P_{tx}$ is the transmission power of each transmitter, $h_{ji}$ and $d_{ji}$ are channel coefficients and distance from the transmitter of link $j$ to the receiver of link $i$ respectively. Assuming $h_{ji}$ is Rayleigh fading and modeling it as an independent random variable across links and time slots with $h_{ji} \sim \mathrm{exp}(1)$. $\alpha$ is the path-loss exponent and $\sigma^2$ denotes the power of additive white Gaussian noise (AWGN).
\subsection{Conditional success probability} \label{mu}
We consider that a transmission is successful when the SINR at the corresponding receiver is greater than a threshold value $\beta_i$. The transmission success probability $\mu_i(k)$ at link $i$ during time slot $k$ is given by
\begin{equation}
	\mu_i(k) \triangleq \mathbb{P}[\mathrm{SINR}_i(k)>\beta_i] \label{1} .
\end{equation}

From \eqref{1}, it is clear that $\mu_i(k)$ depends on four parts: scheduling decision,  buffer state, Rayleigh fading, and the spatial locations of the interfering transmitters,  while buffer state is with respect to the evolution of the system and the past scheduling decisions. Given a particular scheduling decision $\textbf{A}(k)$ and the buffer state $\textbf{N}(k)=\{n_1(k), n_2(k),\dots, n_N(k) \}$, the randomicity of the transmission success probability at link $i$ is only determined by Rayleigh fading. Thus, we can derive the conditional success probability for activated links as follows
\begin{align}
		&\mu_i(k|\textbf{A}(k), \textbf{N}(k))=\mathbb{P}[\mathrm{SINR}_i(k)>\beta_i|\textbf{A}(k),\textbf{N}(k)] \nonumber\\
		=&\mathbb{P}\left[h_{ii}>\frac{\beta_i(\sum_{j\neq{i}}{P_{tx}h_{ji}\lVert{d_{ji}}\rVert ^{-\alpha}a_j(k)n_j(k)}+\sigma^2)}{P_{tx}\lVert{d_{ii}}\rVert^{-\alpha}}\bigg|\textbf{A}(k),\textbf{N}(k)\right] \nonumber\\
		=&\mathbb{E}\left[e^{\frac{-\beta_i\sigma^2}{P_{tx}\lVert{d_{ii}}\rVert^{-\alpha}}}\prod_{j\neq{i}}{\mathrm{exp}\left(-\frac{h_{ji}a_j(k)n_j(k)\beta_i\lVert{d_{ji}}\rVert^{-\alpha}}{\lVert{d_{ii}}\rVert^{-\alpha}}\right)}\bigg|\textbf{A}(k),\textbf{N}(k)\right] \nonumber\\
		=&\rho_i\prod_{j\neq{i}}{\frac{1}{1+a_j(k)n_j(k)/D_{ji}}}  \label{2},
\end{align}
where $\rho_i=\mathrm{exp}\left(-{\beta_i\sigma^2}/{P_{tx}\lVert{d_{ii}}\rVert^{-\alpha}}\right)$ represents the impact of AWGN on the success probability, and $D_{ji}={{\lVert{d_{ii}}\rVert^{-\alpha}}/\beta_i\lVert{d_{ji}}\rVert^{-\alpha}}$. The last equality derives the expectation of $h_{ji}$ by noticing that Rayleigh fading of different channels are independent and identically distributed (i.i.d.).

We can see from \eqref{2} that the conditional success probability is determined by the scheduling decisions  $\textbf{A}(k)$ and the buffer state $\textbf{N}(k)$, where $\textbf{N}(k)$ is determined by the past system evolution of all links. This results in an extremely complex spatio-temporal coupling correlation among the $\mu_i(k)$.
The scheduling decision of any link will affect the transmission success probability of all other links, and then affect the buffer states in the subsequent time slot, which in turn affects the evolution of this link. It is hard to precisely evaluate the impact of specific decisions on the system and the subsequent evolution of the network dynamics.  Therefore, we utilize the stationary randomized policy in Section \ref{policy} to schedule D2D links, which is similar to the random access network and slotted ALOHA protocol\cite{yang2021spatiotemporal}, to simplify the subsequent analysis.

\subsection{Age of Information}
The Age of Information represents the age of the latest information utilized by the receiver. The instantaneous AoI will be updated when a new packet is successfully transmitted. Let $b_i(k)$ indicate whether the transmission is successful. The value of $b_i(k)$ is subject to $a_i(k)$,  $n_i(k)$ and $\mu_i(k)$. If the link $i$ is scheduled to transmit at the time slot $k$  and the buffer is non-empty, i.e., $a_i(k)=1$ and $n_i(k)=1$, then $b_i(k)=1$ with probability $\mu_i(k|\textbf{A}(k),\textbf{N}(k))$ and $b_i(k)=0$ with probability $1 - \mu_i(k|\textbf{A}(k),\textbf{N}(k))$. As for the link $i$ not activated to transmit, i.e.,  $a_i(k)n_i(k)=0$, then $b_i(k)=0$. The conditional expectation of $b_i(k)$ can be obtained as $\mathbb{E}\{b_i(k)|\textbf{A}(k)\}=a_i(k)\mu_i(k|\textbf{A}(k),\textbf{N}(k))$. Substituting \eqref{2} into this equation, we can have
\begin{equation}
	 \mathbb{E}\{b_i(k)|\textbf{A}(k),\textbf{N}(k)\}=\rho_ia_i(k)n_i(k)\prod_{j\neq{i}}{\frac{1}{1+a_j(k)n_j(k)/D_{ji}}}. \label{e3}
\end{equation}

\begin{figure}[t]
	\centering
	\includegraphics[width=10cm]{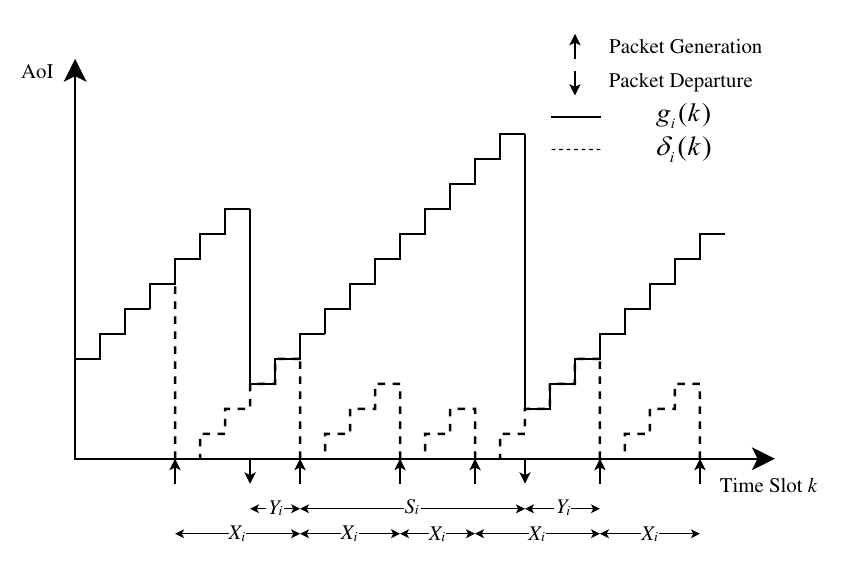}
	\caption{An example of the AoI evolution of link $i$. Solid and dashed lines represent the AoI from the view of receiver and the up-to-date packet respectively. }
	\label{fig-age}
\end{figure}

Let $g_i(k)$ be the instantaneous AoI of the receiver at link $i$ at the time slot $k$. $g_i(k)$ will be updated at the beginning of each slot based on the transmission condition of the previous slot and the buffer state after the potential packet arrival. Denote $\delta_i(k)$ as the instantaneous AoI from the perspective of the up-to-data packet generated in link $i$, which evolves as follows:
\begin{align}
	\delta_i(k)=\begin{cases} 0 &, \mathrm{if}\  e_i(k)=1 \\ \delta_i(k-1)+1 &, \mathrm{if}\ e_i(k)=0, \end{cases} \label{n1}
\end{align}
where $e_i(k)$ denotes the packet generation process with $\mathbb{P}[e_i(k)=1] = \xi$. If the the receiver successfully decodes the packet transmitted at the time slot $k$, i.e., $b_i(k)=1$, the instantaneous AoI will be reset to $\delta_i(k) + 1$ at the next time slot $k+1$. In contrast, if the link $i$ does not complete a successful transmission, then $g_i(k+1)=g_i(k)+1$. Therefore, the evolution of $g_i(k)$ follows
\begin{align}
	g_i(k+1)=\begin{cases} \delta_i(k) + 1 &, \mathrm{if}\  b_i(k)=1 \\ g_i(k)+1 &, \mathrm{otherwise}. \end{cases} \label{e4}
\end{align}
Note that $g_i(k)$ updates at the next slot while $\delta_(k)$ updates at the generation slot. Fig. \ref{fig-age} depicts an example of the age evolution in link $i$. $X_i$ is the interval of the packet generation. $Y_i$ and $S_i$ are queue empty time and packet transmission time respectively.

We use the average AoI, denoted by $\Delta_i = \lim_{K \to\infty}\mathbb{E}\left[\sum_{k=1}^{K}g_i(k)\right]/K$,  to characterize the overall freshness of the time-sensitive information that the receiver at link $i$ cares about. The expectation in the equation is with respect to the randomness of the scheduling decisions and the channel during the whole $K$ time slots. To evaluate the freshness of all $N$ links in the whole D2D network, we use the overall average AoI to reflect fairness, given by
\begin{equation}
	 \varDelta^{\mathrm{Ave}} = \frac{1}{N}\mathbb{E}\left[\sum_{i=1}^{N}\Delta_i\right] =  \lim_{K \to\infty}\frac{1}{KN}\mathbb{E}\left[\sum_{i=1}^{N}\sum_{k=1}^{K}g_i(k)\right] \label{e6}.
\end{equation}
\subsection{Throughput}
We assume that the packets delivered by each link are standardized into uniform packet size\cite{5061954}. The throughput is normalized to $1$ for each successful transmission and $0$ for the failure delivery trial, which is equivalent to the indicator $b_i(k)$ defined above. We denote the long-term average throughput of link $i$ as $T_i = \lim_{K \to\infty}\mathbb{E}\left[\sum_{k=1}^{K}b_i(k)\right]/K$ and the overall average throughput of all links as follows:

\begin{equation}
	T^{\mathrm{Ave}} = \frac{1}{N}\mathbb{E}\left[\sum_{i=1}^{N}T_i\right] = \lim_{K \to\infty}\frac{1}{KN}\mathbb{E}\left[\sum_{i=1}^{N}\sum_{k=1}^{K}T_i(k)\right] \label{e66}.
\end{equation}

\subsection{Scheduling Policy and Optimization Objective} \label{policy}
In this paper, We focus on the design of the scheduling policy for the decision of $\textbf{A}(k)$ to minimize the overall average AoI $\Delta^{Ave}$ and maximize the overall average throughput $T^{Ave}$ simultaneously, which is a multi-objective optimization problem.
Due to the coupling spatio-temporal correlation of transmission success probability in the D2D network, making scheduling decisions needs to collect and process $N^2$ cross-channel state information and $N$ instantaneous AoI and buffer states generated by $N$ direct links. For large-scale networks, real-time dynamic decision-making can cause enormous communication and computation overhead, which is expensive for both centralized and distributed scheduling in time and power \cite{8340813}. Besides, the instantaneous scheduling decision will have an impact difficult to evaluate on the subsequent state of the system due to this coupling correlation.

Therefore, we employ a stationary randomized policy to schedule the D2D links by considering a static problem. This policy has been discussed in some other scheduling scenarios \cite{Scheduling,8943134} and its related practical implements are random access networks and slotted ALOHA. This policy means that, at each time slot, the link $i$ has a time-invariant activation probability $p_i$ independent of other links. The stationary randomized policy is denote by $\pi = \{p_1, p_2,\dots, p_N\}$ and $\Pi$ is the class of all these policies, $\pi\in\Pi$.
Then the scheduling decisions $\textbf{A}(k)$ under this policy turn into random variables with Bernoulli distribution, i.e., $a_i(k)=1$ with probability $p_i$ across all time slots while $a_i(k)=0$ with probability $1-p_i$.
Note that $\pi$ is a static scheduling policy and does not need to collect and process system information in each time slot, although it sacrifices some potential performance.

We define this multi-objective optimization problem under the stationary randomized policy as:
\begin{subequations}\label{151}
	\begin{align}
		&\min_{\pi\in\Pi} \ \  \Delta^{Ave}  \\
		&\max_{\pi\in\Pi} \ \  T^{Ave} \ \ \ \label{15a} \\
		&\ \mathrm{s.t.}\ \   0<p_i\leq1, \ \forall i \in \{1,2,\dots,N\},\label{15b} 		
	\end{align}
\end{subequations}

The goal of solving this problem is to obtain all the Pareto optimal vectors $\{\Delta^{Ave}, T^{Ave}\}$, which is defined as the vectors that no other vectors can have a lower $\Delta^{Ave}$ and a higher $T^{Ave}$ simultaneously. The set of all the Pareto optimal vectors is the Pareto front (PF), which is equivalent to the trade-off curve of information freshness and throughput. To obtain the PF, we convert the multi-objective optimization problem into a scalar optimization problem by the weighted sum approach\cite{MOEA} which considers the affine combination of $\Delta^{Ave}$ and $1/T^{Ave}$ as the objective function. Thus, the scalar optimization problem can be obtained as
\begin{subequations}\label{obj}
	\begin{align}
		\min_{\pi\in\Pi}\ \ &\mathcal{P}(\lambda)=\lambda\Delta^{Ave}+(1-\lambda)\frac{1}{T^{Ave}} \label{eq15a} \\
		\mathrm{s.t.}\ \  &0<p_i\leq1, \ \forall i \in \{1,2,\dots,N\},\label{eq15b}
	\end{align}
\end{subequations}
where $\lambda$ and $1-\lambda$ are the weights for $\Delta^{Ave}$ and $1/T^{Ave}$ respectively, $\lambda\in[0,1]$.
We can obtain the scheduling policy parameters for the D2D network under different importance assignments between information freshness and throughput by solving this problem with different $\lambda$. For each $\lambda\in[0,1]$, the optimal solution of the scalar optimization problem forms a Pareto optimal vector.

It is worth noting that although this is a static problem, its optimization is still difficult. Specifically, the spatio-temporal correlation of link states makes it tough to obtain the precise closed-form expression of this problem, and even if the expression is obtained, the problem is still non-convex. In section \ref{problem} and \ref{dl}, we respectively derive an implicit expression by introducing the mean-field assumption and solving this problem with a deep learning algorithm.

\section{Problem formulation} \label{problem}
In this section, we derive the transmission success probability, buffer non-empty probability, overall average AoI, and overall average throughput under the mean-field assumption and formulate the optimization problem. We also remark a special case when the packet arrival rate $\xi = 1$ where the buffers are fixed stuffed and introduce an iterative approach to solve this special case for comparison.


\subsection{Queue state derivation}
With the employment of stationary randomized policy, the evolution and dynamics of the network will be determined by the randomicity of policy decisions. However, as mentioned in section \ref{mu}, the interaction between links causes the spatio-temporal correlation of node states, which makes it difficult to carry out statistical analysis.
We introduce the mean-field assumption \cite{15561801, yang2021spatiotemporal} to enable the subsequent derivation. This assumption grants an interacting particle system can utilize the time average state to represent its time-varying properties under mutual correlation. Specifically, we assume the buffer states of each link are i.i.d. over other links and time-varying according to its steady-state non-empty probabilities\cite{8688635}, denoted by $\{\nu_i\}_{i=1}^N = \lim_{K \to\infty}\sum_{k=1}^{K}n_i(k)/K \in (0,1]$. Accordingly, the evolution of the queue state of each link can be regarded as independent of each other, which allows us to analyze the transmission process of each link separately.

\emph{1) Transmission success probability:}Based on the non-empty probabilities, we can derive the expectation of the transmission success probability under the steady-state as follows
\begin{align}
	\mu_i\triangleq\lim_{k\to\infty}\mathbb{E}[\mu_i(k)|\pi]=&\lim_{k\to\infty}\mathbb{E}\left[\rho_i\prod_{j\neq{i}}{\frac{1}{1+a_j(k)n_j(k)/D_{ji}}}\bigg|\pi\right] \nonumber \\
	=&\rho_i\prod_{j\neq{i}}{\left(1-\frac{p_j\nu_j}{1+D_{ji}}\right)}, \label{mupi}
\end{align}
where the equality holds by taking the expectation of the i.i.d. random variable $a_j(k)$ and $n_j(k)$. Therefore, under the stationary randomized policy and mean-field assumption, the packet transmission process and transmission time of link $i$ can be considered as Bernoulli process and geometric distribution subject to parameter $p_i\mu_i$, respectively.

\emph{2) Buffer non-empty probability:}To obtain the transmission success probability for a particular policy parameter $\pi$, we still need to calculate the buffer non-empty probability $\{\nu_i\}_{i=1}^{N}$. Benefiting from the independence resulting from the mean-field assumption, we can derive the buffer non-empty probability seperately from the respective evolution. As depicted in Fig. \ref{fig-age}, the time ratio of the packet transmission time $S_i$ to the interval time of two success transmission $Y_i + S_i$  is the buffer non-empty probabilitys: $\nu_i = \mathbb{E}[S_i]/\mathbb{E}[Y_i+S_i]$.
Note that $Y_i$ is the buffer empty time which is equivalent to the time slots number without packet generation consecutively, i.e., $\mathbb{E}[Y_i]= \mathbb{E}[X_i] - 1$. $X_i$ and $S_i$ are both geometric distribution subject to parameter $\xi$ and $p_i\mu_i$ respectively, and $\nu_i$ can be derived as follows:
\begin{align}
	\nu_i = \frac{\frac{1}{p_i\mu_i}}{\frac{1}{\xi} + \frac{1}{p_i\mu_i}} = \frac{\xi}{\xi + (1 - \xi)p_i\mu_i} \label{nu}
\end{align}

\emph{3) Age of information and throughput:} Based on \eqref{mupi} and \eqref{nu}, we can derive the expressions of the two metrics we adopt in this paper as follows:
\begin{align}
	\varDelta^{\mathrm{Ave}}& = \frac{1}{N}\sum_{i=1}^{N}\left[\frac{1}{\xi} + \frac{1}{p_i\mu_i}-1\right]  \label{aoi_all}\\
	T^{\mathrm{Ave}}& = \frac{1}{N}\sum_{i=1}^{N}\left[\frac{\xi p_i \mu_i}{\xi + (1 - \xi)p_i\mu_i}\right]. \label{thr_all}
\end{align}
\begin{proof}
See Appendix \ref{app1}.	
\end{proof}

\subsection{Problem formulation}
Substituting \eqref{mupi}-\eqref{thr_all} into \eqref{obj}, we formulate the optimization problem as follows:
\begin{subequations}\label{problem2}
	\begin{align}
		\min_{\pi\in\Pi}\ \ &\mathcal{P}(\lambda)=\frac{\lambda}{N}\sum_{i=1}^{N}\left[\frac{1}{\xi} + \frac{1}{p_i\mu_i}-1\right]+\frac{1-\lambda}{\frac{1}{N}\sum_{i=1}^{N}\left[\frac{\xi p_i \mu_i}{\xi + (1 - \xi)p_i\mu_i}\right]} \label{eq15a} \\
		\mathrm{s.t.}\ \  & \eqref{mupi}, \  \eqref{nu}, \\
		& 0<p_i\leq1, \ \forall i \in \{1,2,\dots,N\}.\label{eq15b}
	\end{align}
\end{subequations}

It is worth noting that the network queue dynamic will gradually converge to the steady-state according to the specific policy parameters $\pi$, which implies that $\{\nu_i\}_{i=1}^{N}$ and $\{\mu_i\}_{i=1}^{N}$ are both functions of $\{p_i\}_{i=1}^{N}$. Nonetheless, the nesting of $\{\nu_i\}_{i=1}^{N}$ and $\{\mu_i\}_{i=1}^{N}$ makes it hard to derive their explicit expression. Consequently, the influence of the optimization variables $\{p_i\}_{i=1}^{N}$ on $\mathcal{P}(\lambda)$ is very difficult to analyze. To solve this problem, we propose a neural network structure to constitute the mapping from the D2D network parameters to the optimal solution of \eqref{problem2} and train the neural network unsupervised in Section \ref{dl}.
	
It can be seen from \eqref{a_i} and \eqref{thr_i} that the long-term throughput is the reciprocal of the average AoI for a single link where the average AoI decreases as the throughput increases. Obviously, there is no trade-off between their optimizations under LCFS-PR Geo/Geo/1 queue for a single link, and they can reach the optimal point simultaneously when $p_i\mu_i\to \sup\{p_i\mu_i\}$. Nonetheless, when it comes to the scheduling for the whole network performance, the optimal point $\sup\{p_i\mu_i\}$ of each link $i$ cannot be achieved simultaneously. The trade-off between $\Delta^{Ave}$ and $T^{Ave}$ mainly depends on the network resources competition and coupling correlation between links. Intuitively, the $T^{Ave}$ maximization drives links with better channel condition to achieve higher throughput, while links with worse channel is on the contrary. Due to $T_i \in [0,1]$, the average AoI of links with the bad channel, which is the reciprocal of $T_i$, will increase sharply and result in poor $\Delta^{Ave}$ performance. More specific details on the freshness-throughput trade-off are illustrated in Section \ref{numerical}.

\subsection{Special case and an iterative approach} \label{sp}
We consider a special case for problem \eqref{problem2} with the packet generating rate $\xi =1$. In this case, each buffer will have the freshest packets available for transmission at any time slot, which means that the buffer non-empty probability is $1$. Consequently, the transmission success probability $\mu_i$ at each time slot can be precisely derived by an explicit expression $\mu_i=\rho_i\prod_{j\neq{i}}{\left(1-\frac{p_j}{1+D_{ji}}\right)}$ without the mean-field assumption owing to the Rayleigh fading and link activation probability being i.i.d.. The optimization problem can also be obtained in an explicit form as follows:
\begin{subequations}\label{problem3}
	\begin{align}
		\min_{\pi\in\Pi}\ \ &\hat{\mathcal{P}}(\lambda)=\frac{\lambda}{N}\sum_{i=1}^{N} \frac{1}{p_i\rho_i\prod_{j\neq{i}}{\left(1-\frac{p_j}{1+D_{ji}}\right)}}+\frac{1-\lambda}{\frac{1}{N}\sum_{i=1}^{N}p_i\rho_i\prod_{j\neq{i}}{\left(1-\frac{p_j}{1+D_{ji}}\right)}} \label{eq151} \\
		\mathrm{s.t.}\ \ & 0<p_i\leq1, \ \forall i \in \{1,2,\dots,N\}.\label{eq152}
	\end{align}
\end{subequations}

Before introducing the deep learning algorithm, we first propose an iterative approach to acquire a great local optimal solution for problem \eqref{problem3}. We utilize this approach as a comparison with the proposed deep learning algorithm in Section \ref{dl}. The flow of this algorithm is to optimize the activation probability of each link separately by fixing other links. An iteration consists of this univariate optimization step for all links once. After several iterations, $\hat{\mathcal{P}}(\lambda)$ converges to a local optimal solution.

Specifically, \eqref{eq151} is a non-negative linear combination of a series of convex functions when only optimizing the activation probability of a single link with other links fixed, which can be solved by dichotomy or by a commercial convex optimization solver such as CVX. In each step of the iteration, the activation probabilities of the other links utilize the latest obtained $p_i$ in the previous step, and $\hat{\mathcal{P}}(\lambda)$ is non-increasing. Consequently, $\hat{\mathcal{P}}(\lambda)$ can finally converge to a local optimal value.
This approach is difficult to apply because it requires a centralized collection of all $N^2$ channel gain information and a large amount of computation. We only use it as a comparison for the deep learning algorithm proposed in the next Section.

\section{Deep Learning Algorithm} \label{dl}

\textcolor{blue}{When utilizing a traditional algorithm to solve link scheduling problems, the path-loss of each channel must be obtained first. However, for the channel of $N$ links interfering with each other under a D2D fashion, the  path-loss matrix consists of $N^2$ elements to be processed and therefore the computational complexity of the traditional approach will be at least $O(N^2)$. Meanwhile, computation and collection of path-loss are also expensive and resource-consuming.}

Inspired by \cite{8664604}, where the author utilizes a spatial convolutional neural network to learn the mapping from the $N$ geographic location information (GLI) to the optimal schedule of the sum-rate maximization problem, we propose a neural network structure to obtain the AoI and throughput trade-off curve with a similar idea. This technique greatly reduces the scale of information required to make decisions, and GLI can be obtained easily by utilizing GPS.
\textcolor{blue}{In fact, the $N^2$ CSI required to schedule links are determined by Rayleigh fading and path loss in our model, where the Rayleigh fading is captured by the random variable distribution model and path loss which can be calculated by the GLI of each link.
Thus, the mapping from GLI to the optimal solution can be learned by the neural network without the estimation of CSI. }

Different from \cite{8664604}, the objective function of our AoI and throughput jointly scheduling problem can not be expressed explicitly, which is hard to solve directly by gradient descent. We design a different training approach for this problem to make the neural network work. Besides, since the packet generation process parameter and the AoI-throughput trade-off weight are varying, we train the neural network to learn the mapping of any \{GLI, $\xi$, $\lambda$\} combinations to the best policy parameters $\pi$. The specific network structure and working fashion are introduced in the following part of this section, which is also depicted in Fig. \ref{fig2}.

\subsection{Input pre-processing}
We assume that the D2D plane is a square with a side length of $L$ and the GLI is defined as a set of vectors $\{(x_i^\mathrm{tx},y_i^\mathrm{tx}),(x_i^\mathrm{rx},y_i^\mathrm{rx})\}_{i=1}^N$ representing the geographic locations coordinates of the transmitters and receivers, with the coordinate values ranging from  $0$ to $L$.

To input the GLI into the neural network, we preprocess the GLI into matrices to build more appropriate forms. Specifically, we divide the D2D plane into square grids with the size of $M\times M$, which is also the size of the device  matrices transformed from the GLI. Denote $G^{i}_{TX}$ and $G^{i}_{RX}$ as the transmitter matrix and the receiver matrix of link $i$ respectively, where both the two matrices are one-hot matrices to imply the location of the devices and defined as follows
\begin{align}
	G^{i}_{TX}(x,y)=\begin{cases} 1 &, \mathrm{if}\  (x,y)=\lceil (x_i^\mathrm{tx},y_i^\mathrm{tx} )* M / L \rceil \\ 0 &, \mathrm{otherwise}. \end{cases} \label{onehot}
\end{align}

It is worth noting that unactivated links have no effect on the network, and the activation probability $p_i$ indicates how frequently the link $i$ is activated. We define the network GLI matrix under a particular scheduling decision by
\begin{align}
	G^{TX}=\sum_{i=i}^{N}p_iG^{i}_{TX}(x,y), \label{gtx2}
\end{align}
which is the same for the receiver matrix. Activation probability $\{p_i\}_{i=1}^N$ is used to construct the matrix to reflect the effect of activation probability on other devices, e.g., transmitters with higher $p_i$ cause more interference to other links. Therefore, $G_{TX}$ and $G_{RX}$ can be regarded as a feature matrix that contains most of the GLI information required to solve \eqref{problem2}. It is worth noting that the preprocess of GLI constructs a fixed input form regardless of the size of $N$ and forms matrices suitable for passing through the convolution layer.
\addtolength{\topmargin}{0.36cm}

\begin{figure}[t]
	\centering
	\includegraphics[width=14cm]{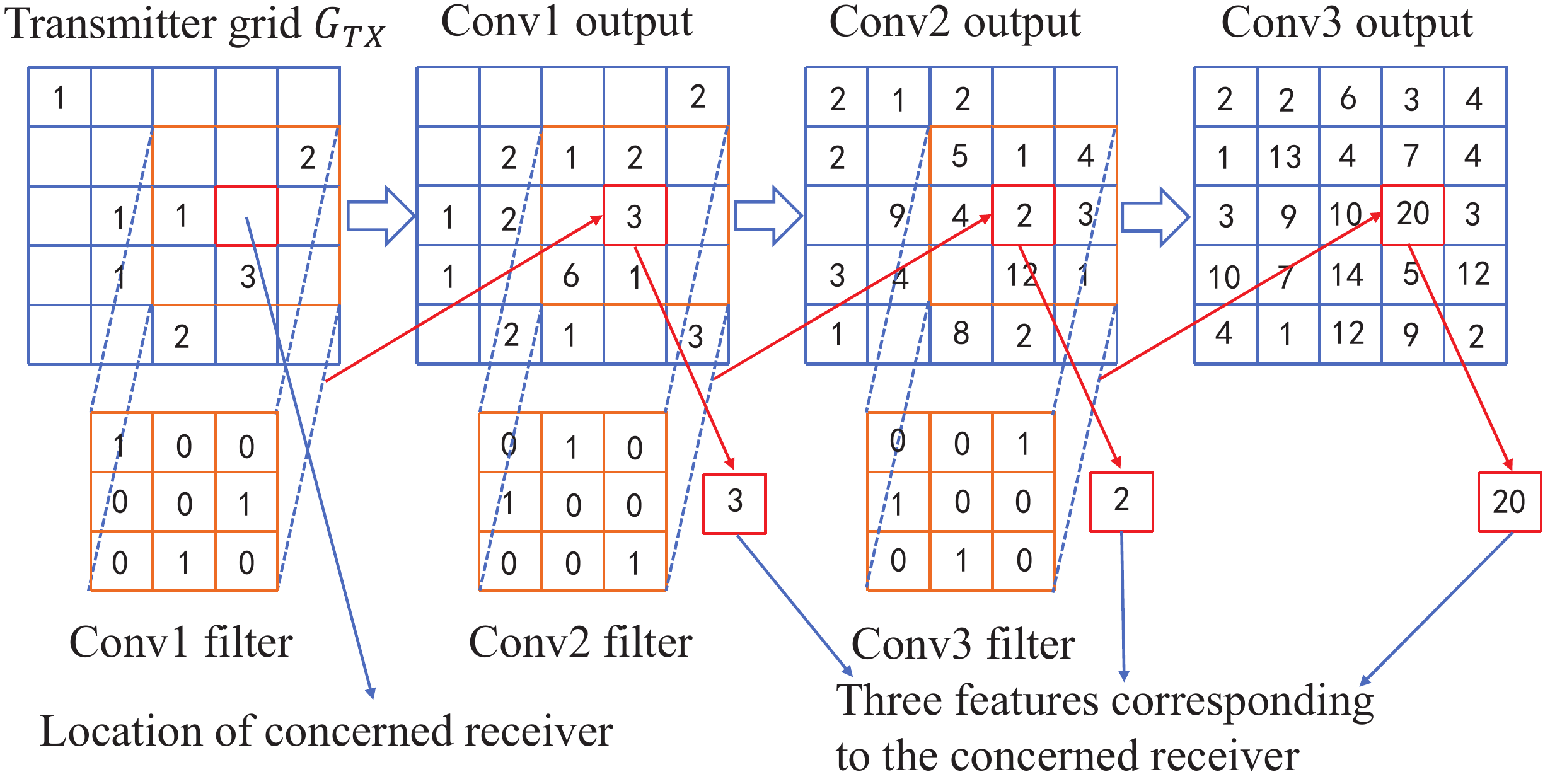}
	\caption{An example of a transmitter grid passing through convolution stage. In every output matrix, each link extracts a number from the red cell, where is the location of the receiver for this link, as a feature of the GLI of the transmitters nearby.}
	\label{fig1}
\end{figure}

\subsection{Convolution Stage}
Different from \cite{8664604} using a single convolution layer to simulate the total interference caused by other devices, we regard the convolution layer as a way to extract the feature from the GLI of transmitters around each receiver or receivers around each transmitter. Experiments show that our structure can output solutions with similar performance compared with the reference network when deployed to solve \eqref{problem2}, but our training is more stable and the convergence rate is greatly improved.

We use three connected convolution layers to extract GLI features, as depicted in an example in Fig. \ref{fig1}, and output a set of features for each link after each convolution layer. Fig. \ref{fig1} illustrates the processing of a transmitter grid $G_{TX}$ using arbitrarily set data. The transmitter grid passes through three convolution layers in turn with padding and outputs three matrices of the same size. Each element in the output matrix is obtained by a single convolution centered at the same index of the input matrix with the convolution filter. Consider this index as the location of a receiver, and convolution can be considered as extracting feature from all transmitters within the range of convolution filter size near this receiver. Assume that the sizes of the three convolution filters are $n_1,n_2,n_3$, respectively. Three connected convolution layers enlarge the scope of feature extraction in stages, and each element of the last output matrix contains the original grid information of $n_1+n_2+n_3-2$ size.

The convolution stage operates symmetrically on the receiver grid $G_{RX}$ with the same convolution filter to extract GLI features of the receivers around a particular transmitter. $G_{TX}$ and $G_{RX}$ pass through the convolution stage in parallel and generate three output matrices respectively. Each link in the D2D plane extracts six features in total from these matrices by the index of its own receiver or transmitter.
\begin{figure}[t]
	\centering
	\includegraphics[width=10cm]{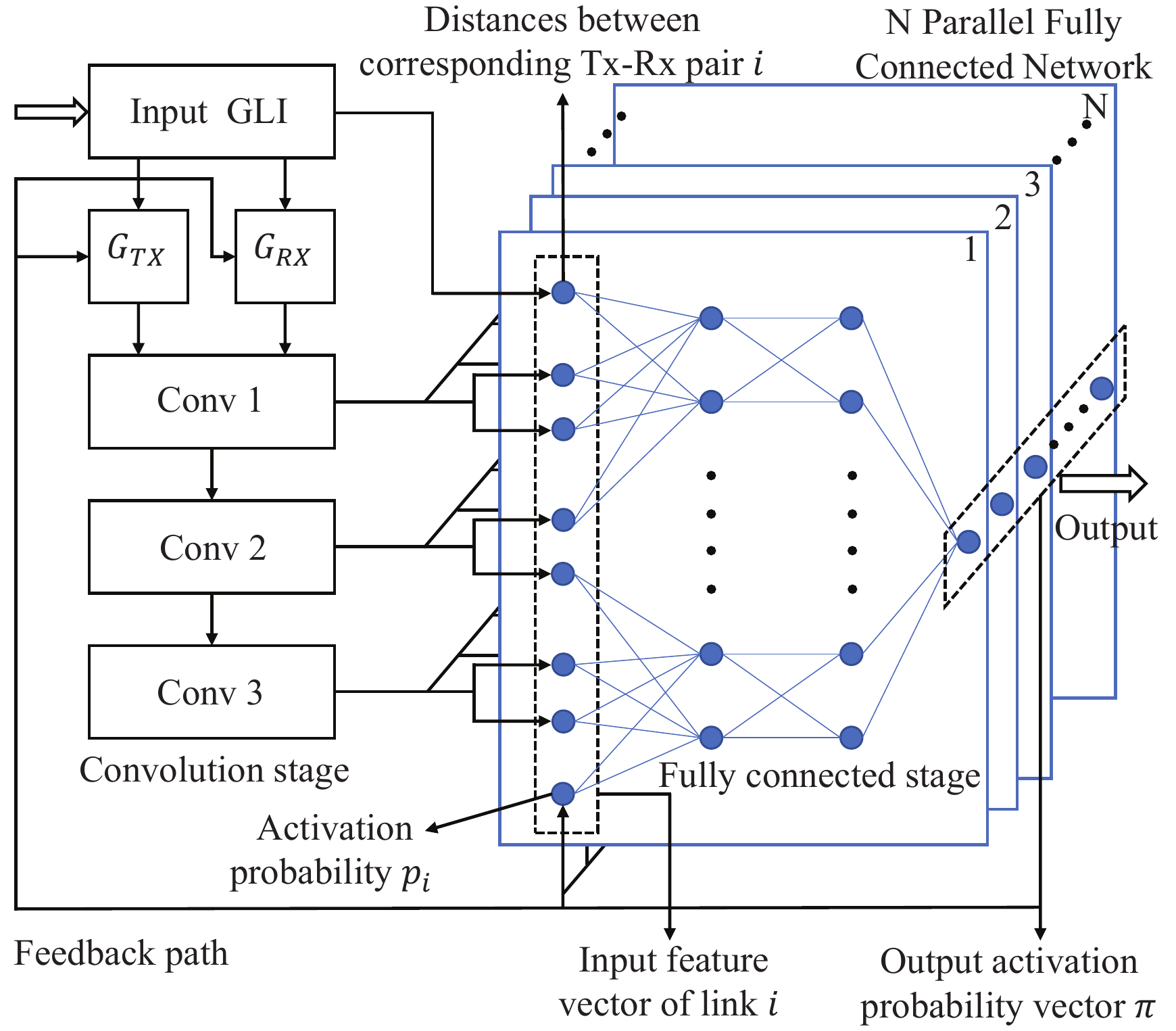}
	\caption{The overall structure of the neural network. }
	\label{fig2}
\end{figure}

\subsection{Fully Connected Stage}
Similar to the general neural network structure, we input the intermediate features into a fully connected network to generate the output with the desired form. The fully connected network consists of two hidden layers whose activation function is the rectified linear unit and finally outputs the activation probability $p_i\in(0,1)$ after passing through a sigmoid function. For a D2D layout with $N$ links, the feature vectors of all links pass through the fully connected layer respectively and output a set of activation probability vectors $\pi$ at the end.

In addition to the six GLI features from the previous convolution stage into the fully connected stage, we also input four other features. Specifically, the first is the activation probability $p_i$ output from the previous iteration, which is used to enhance the feedback. The second is the distance between the receiver and the transmitter of this link, denoted by  $d_i^{TR}=\Vert(x_i^\mathrm{tx}-x_i^\mathrm{rx},y_i^\mathrm{tx}-y_i^\mathrm{rx})\Vert_2$. This is due to the feature vector output from the convolution stage can only characterize the GLI around the transmitter and receiver of the concerned link and the relationship between the transceiver of the link itself is indeterminate, which is an important feature that needs the neural network to learn. The third and the fourth are $\xi$ and $\lambda$ respectively, which are used to identify the different requirements of the output solution. With these two features as inputs, the neural network does not need to be retrained for different packet generation rates $\xi$ and AoI-throughput trade-off weight $\lambda$.

\subsection{Iterative fashion}
The entire network structure is constituted by connecting the parts introduced above together through a feedback loop. The network works in the form of an iterative feedback fashion because a single forward path of the network is not sufficient to learn the mapping from the GLI to the optimal policy parameter. We input the output activation probability vectors $\{p_i\}_{i=1}^N$ into the pre-processing stage to form a new GLI with \eqref{gtx2}. $\{p_i\}_{i=1}^N$ will also be feedback to the fully connected stage in the next iteration as a part of the feature vectors. In this way, each iteration actually completes an improvement on the scheduling policy parameters $\{p_i\}_{i=1}^N$ of the previous iteration.

Regardless of how $\{p_i\}_{i=1}^N$ changes during the iteration, the $G^{i}_{TX}(x,y)$ and $G^{i}_{RX}(x,y)$ to construct the $G^{TX}$ and $G^{RX}$ is fixed, which means that no matter which intermediate $\{p_i\}_{i=1}^N$ the iteration experience, the final output $\{p_i\}_{i=1}^N$ should be the same. Owning to the strong approximation ability of the neural network, We find experimentally that the output will converge after several times iteration after the network is well-trained. Consequently, we just set all the initial $\{p_i\}_{i=1}^N$ to $1$ without loss of generality.

\subsection{Network Training}
The optimal solution $\pi^*$ to \eqref{problem2} can be seen as a function of the D2D system parameters. We use $\textbf{C}$ to represent these parameters and we can obtain $\pi^* = \arg \min\{\mathcal{P}(\pi|\textbf{C})\}$, where $\mathcal{P}(\pi|\textbf{C})$ is the objective function of \eqref{problem2} with all the D2D system parameters represented by $\textbf{C}$. The Neural network are essentially a function constituting mappings from inputs to outputs, i.e., $\pi = g(\textbf{w}, \textbf{b},\textbf{C})$, where $\textbf{w}, \textbf{b}$ are neural network parameters to be trained and $g(\cdot)$ represents the structure of this neural network. The training process is to find the optimal $\textbf{w}^*, \textbf{b}^*$ to approximate $\pi^*$, that is, $\textbf{w}^*, \textbf{b}^* = \arg \min\{\mathcal{P}(g(\textbf{w}, \textbf{b},\textbf{C})|\textbf{C})\}$. Consequently, we use the objective function of \eqref{problem2} to be the loss function of the neural network we proposed. This is an unsupervised learning process because for any parameters $\textbf{C}$, we do not need to manually label the corresponding $\pi^*$.

The optimization of neural network parameters is accomplished by gradient descent approach for loss function, which implies that we have to obtain the gradient of $\mathcal{P}(g(\textbf{w}, \textbf{b},\textbf{C})|\textbf{C})$ subject to $\textbf{w}, \textbf{b}$. Generally speaking, the gradients of explicit loss functions can be directly obtained by back-propagation and automatically differentiated using the Tensorflow toolbox. However, the objective function of \eqref{problem2} has only an implicit expression and cannot be back-propagated directly.

By noting that when updating the network parameters $\textbf{w}, \textbf{b}$, only the numerical value of the gradient is required, we have the following derivation:

\begin{align}
	\frac{\partial\mathcal{P}}{\partial \textbf{w}} = \sum_{i=1}^{N} \frac{\partial\mathcal{P}}{\partial p_i}\frac{\partial p_i}{\partial \textbf{w}}, \ \ \ \frac{\partial\mathcal{P}}{\partial \textbf{b}} = \sum_{i=1}^{N} \frac{\partial\mathcal{P}}{\partial p_i}\frac{\partial p_i}{\partial \textbf{b}},
\end{align}
where $p_i$ can be explicit expressed by $\textbf{w}$ and $ \textbf{b}$ and can be back-propagated directly. Our goal is to obtain the numerical value of the gradient $\frac{\partial\mathcal{P}}{\partial p_i}$, which can be derived as follows:
\begin{align}
	&\frac{\partial\mathcal{P}}{\partial p_i} = \lambda\frac{\partial\varDelta^{\mathrm{Ave}}}{\partial p_i} -\frac{1-\lambda}{(T^{\mathrm{Ave}})^2} \frac{\partial T^{\mathrm{Ave}}}{\partial p_i} \\
	&\frac{\partial\varDelta^{\mathrm{Ave}}}{\partial p_i} = -\frac{1}{N}\sum_{j=1}^N \left[\frac{1}{(p_j\mu_j)^2}+\frac{1-\xi}{[\xi + (1-\xi)p_j\mu_j]^2}\right]\left(p_j\frac{\partial\mu_j}{\partial p_i}+\mu_j\frac{\partial p_j}{\partial p_i}\right) \\
	&\frac{\partial T^{\mathrm{Ave}}}{\partial p_i} = -\frac{1}{N}\sum_{j=1}^N \frac{(1-\xi)^2}{[\xi + (1-\xi)p_j\mu_j]^2} \left(p_j\frac{\partial\mu_j}{\partial p_i}+\mu_j\frac{\partial p_j}{\partial p_i}\right),
\end{align}
where $\frac{\partial\mu_j}{\partial p_i} = (P^{-1}Q)_{ji}$,  $(P^{-1}Q)_{ji}$ represents the elements of column $i$ in row $j$ of matrix $P^{-1}Q$. $P$ and $Q$ are obtained as follows (See Appendix \ref{appb} for the derivation):
\begin{align}
	&P_{ji} = \frac{qB_{ij}p_i^2}{(1+qp_i\mu_i)(1+qp_i\mu_i-B_{ij}p_i)}, j\neq i,\ \  P_{jj} = -\frac{1}{\mu_j} \label{Mji}\\
	&Q_{ji} = \frac{B_{ij}}{(1+qp_i\mu_i)(1+qp_i\mu_i-B_{ij}p_i)} \label{Nji}
\end{align}
where $B_{ij}=\frac{1}{1 + D_{ij}}$ and $q = \frac{1-\xi}{\xi}$. The only value have not been derived among the above expressions is $\{\mu_i\}_{i=1}^N$, which can be iteratively obtained by ultilizing \eqref{mupi} and \eqref{nu} alternatively with the initial value $\{\mu_i\}_{i=1}^N=0$. This can be easily proofed to be converged because in each iteration, $\{\mu_i\}_{i=1}^N$ are non-decreasing.

In this way, the implicit expression of problem \eqref{problem2} can be optimized during the neural network training. We randomly generate a large number of i.i.d. D2D layouts as the training set of the network. Specifically, we first generate the locations of $N$ transmitters on the square plane according to a uniform distribution within the plane size $(0,L)$, and then generate a corresponding receiver for each transmitter with the distance $d_i^{TR}$, where $d_i^{TR}\sim U(d_{min},d_{max})$, and the angle is also uniform distributed within $[0,2\pi]$. In actual training, the training set is composed of a randomly generated D2D plane and randomly generated $\xi$ and $\lambda$, which constitutes an infinite number of training samples.

\textcolor{blue}{It is worth noting that the training stage contains the above calculation and naturally requires specific information of $N^2$ path-loss. However, only $N$ GLI is required to obtain the scheduling policy parameters after the neural network well-trained. The specific information of path-loss can be obtained by the computation of GLI.}

\section{Numerical Results} \label{numerical}
In this section, we evaluate the performance of the proposed deep learning approach. Following \cite{FPLinQ,8664604}, we model the interference channel according to the short-range outdoor channel model ITU-1411\cite{ITU}, with $5$ MHz bandwidth at $2.4$ GHz carrier frequency. Antenna height and gain are set to $1.5$ meters and $2.5$ dBi respectively. The transmission power of each device $P_{tx}$ is $40$ dBm and the noise power spectral density is $-169$ dBm/Hz. The decoding threshold $\beta_i$ is set to $0$ dB for all links. We consider D2D scenarios where $N$ links are deployed in a 500 meters by 500 meters square region, and the links are randomly distributed in the way that training sets are generated with $d_{min}=2~\mathrm{m}$ and $d_{max}=65~\mathrm{m}$.

For different numbers of links, we generated $10000$ sets of D2D layouts to train the neural network, and $1000$ sets of D2D layouts for testing respectively. As to the configuration of the neural network, GLI is preprocessed to grid with grid length $M = 125$, the sizes of three convolution filters are all $15 \times 15$ and both two hidden layers of the fully connected stage are set to $30$ neurons. The parameters of the neural network are trained separately for different $N$.

For each training step, $32$ D2D layouts are randomly selected from the training set, and $32$ different $\xi$ and $\lambda$ are randomly generated to construct input data together. $\xi$ and $\lambda$ are subject to uniform distribution range in $(0,1]$. Due to the dynamic range of AoI being too large compared with throughput, we use exponential transformation to change the distribution of the $\lambda$ in loss function evaluation by $\lambda^{loss} = 10^{-5 *(1-\lambda)}$ in order to increase the data amount for small $\lambda$. We use Adam optimizer with the learning rate set to $3$e$-4$.

We compare the performance of the proposed neural network algorithm with the following benchmarks:
\begin{itemize}
	\item \textbf{Iterative Minimization Algorithm:} The local optimal algorithm we proposed in Section \ref{sp} with 20 iterations. This algorithm can only work for packet generation rate $\xi=1$. In the figure legend, we use 'Ite. Min.' to represent this algorithm.
	\item \textbf{Optimal slotted ALOHA Policy \cite{mankar}:} Each link in the network is activated with the same probability to minimize the objective function, i.e., $p_i=p^*,\forall i$. In the figure legend, we use 'ALOHA' to represent this algorithm.
	\item \textbf{Locally adaptive slotted ALOHA \cite{yang2021spatiotemporal}:} The activation probability of each link is decoupled by the mass transportation theorem \cite{8186962}. And then each link can dynamically adjust its activation probability by exchanging system information with other links in its observe window. We set the observe window to infinity to constitute the same condition as other methods. This algorithm can only work for packet generation rate $\lambda=1$. In the figure legend, we use 'Adaptive ALOHA' to represent this algorithm.
\end{itemize}
\subsection{Performance of AoI Minimization}
We first evaluate the performance of the deep learning algorithm in the AoI minimization problem where the AoI weight is set to $\lambda=1$. In the figure legend, we use 'DL Algorithm' to represent this algorithm.

\begin{figure}[t]
	\centering
	\includegraphics[width=10cm]{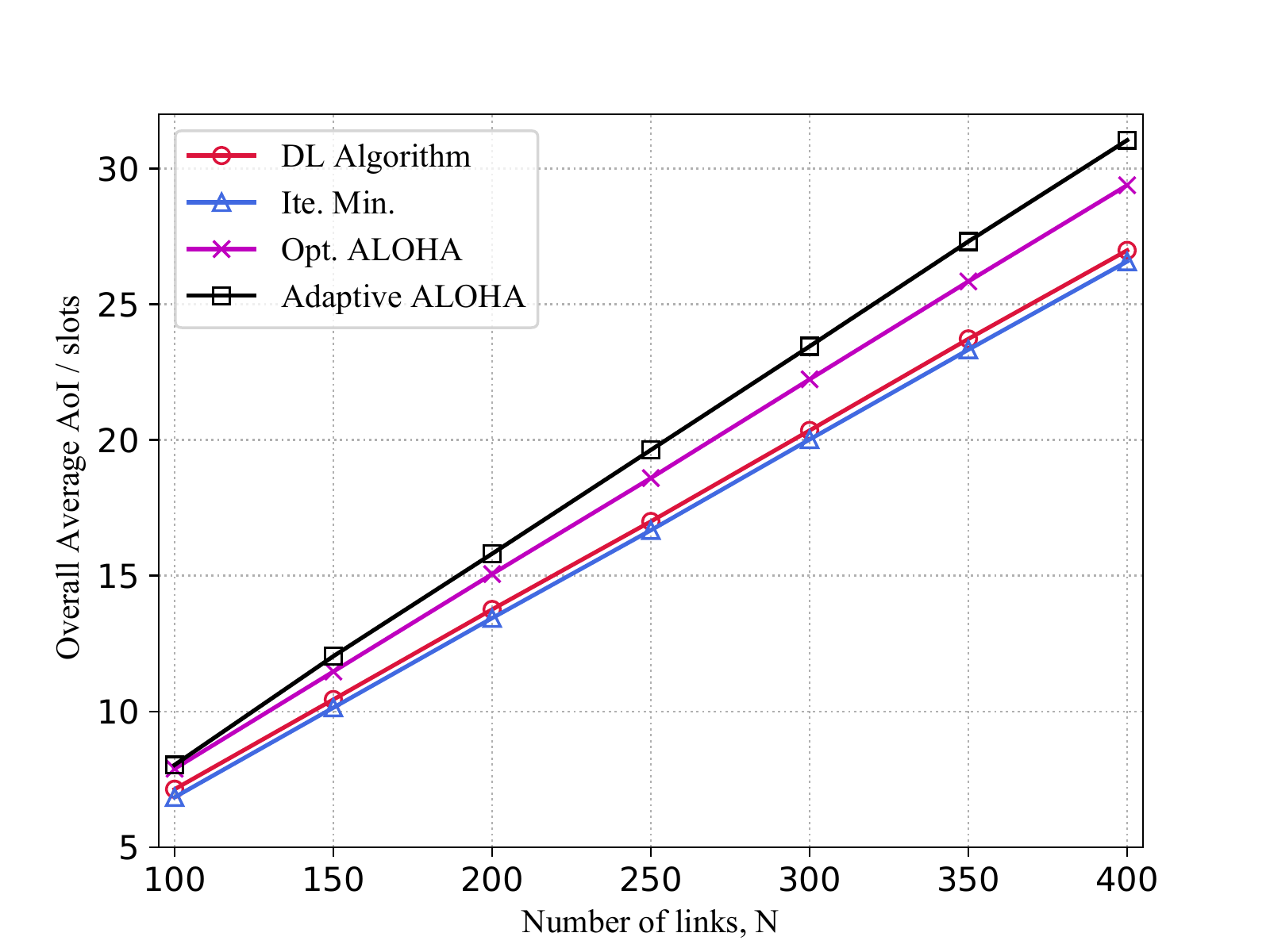}
	\caption{Average $\varDelta^{\mathrm{Ave}}$ performance for $1000$ different D2D layouts as functions of the number of links $N$, where we set $\lambda=1$, $\xi=1$.}
	\label{fig7}
\end{figure}

Fig. \ref{fig7} illustrates the average $\varDelta^{\mathrm{Ave}}$ for $1000$ testing layouts under the stationary randomized scheduling decisions obtained by each method for different numbers of links when $\xi=1$ and $\lambda=1$. We can observe that the proposed deep learning approach achieves very close performance compared with the iterative minimization algorithm while it does not require the computation of path loss and has much less computational complexity, which proves that the mapping from GLI to the scheduling decision can be correctly learned by the proposed neural network structure. What's more, the performance of the deep learning algorithm increases with the growth of $N$. Our algorithm outperforms Adaptive ALOHA and Opt. ALOHA because the former utilizes a lax approximation for decoupling the transmission successful probability between links and the latter cannot adjust the activation probability according to the channel condition differences caused by different link locations.

\begin{figure}[t]
	\centering
	\includegraphics[width=10cm]{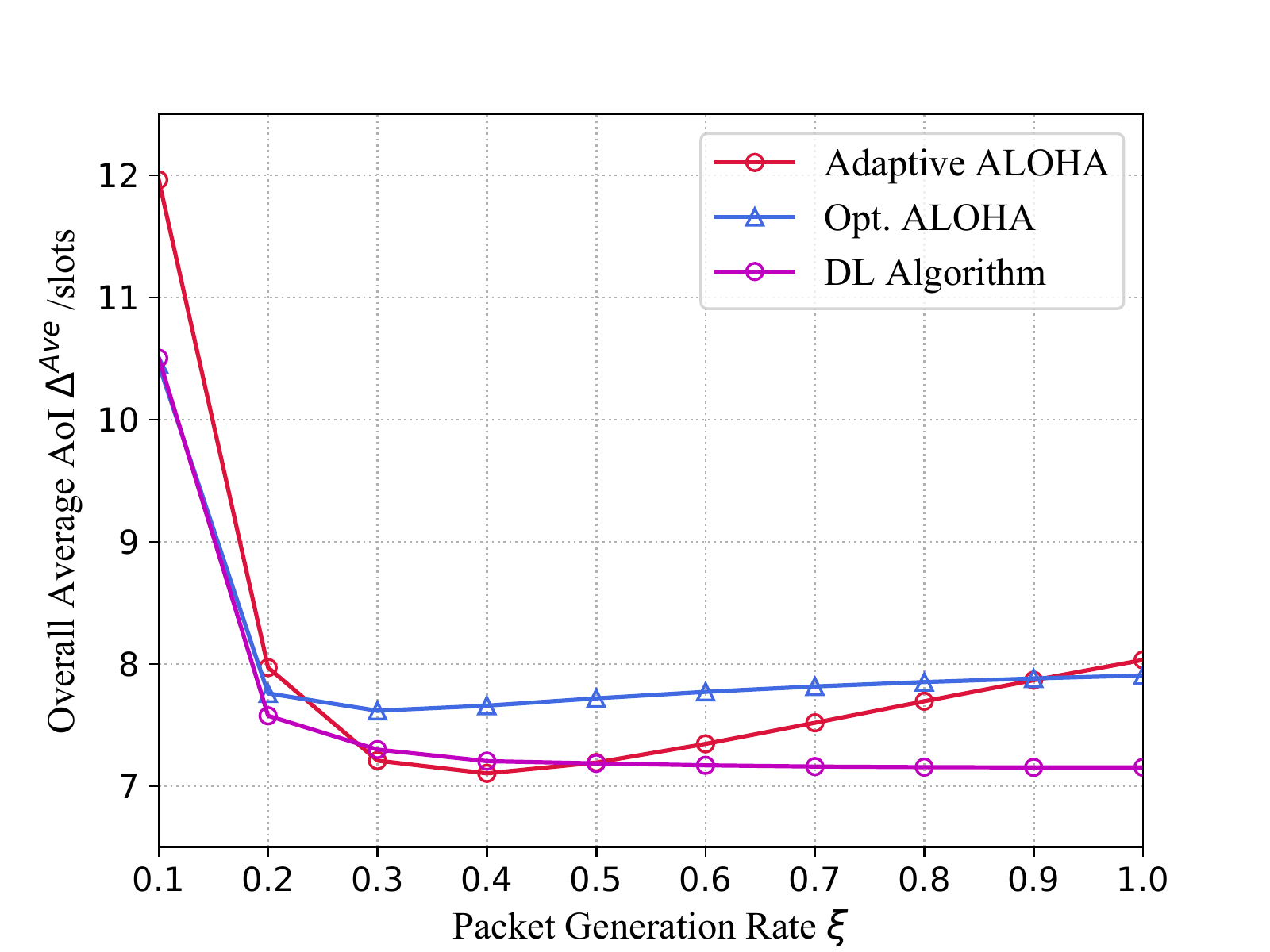}
	\caption{Average $\varDelta^{\mathrm{Ave}}$ performance for $1000$ different D2D layouts as functions of the packet generating rate $\xi$, where we set $\lambda=0$ and $N=100$.}
	\label{fig71}
\end{figure}
\begin{figure}[t]
	\centering
	\includegraphics[width=10cm]{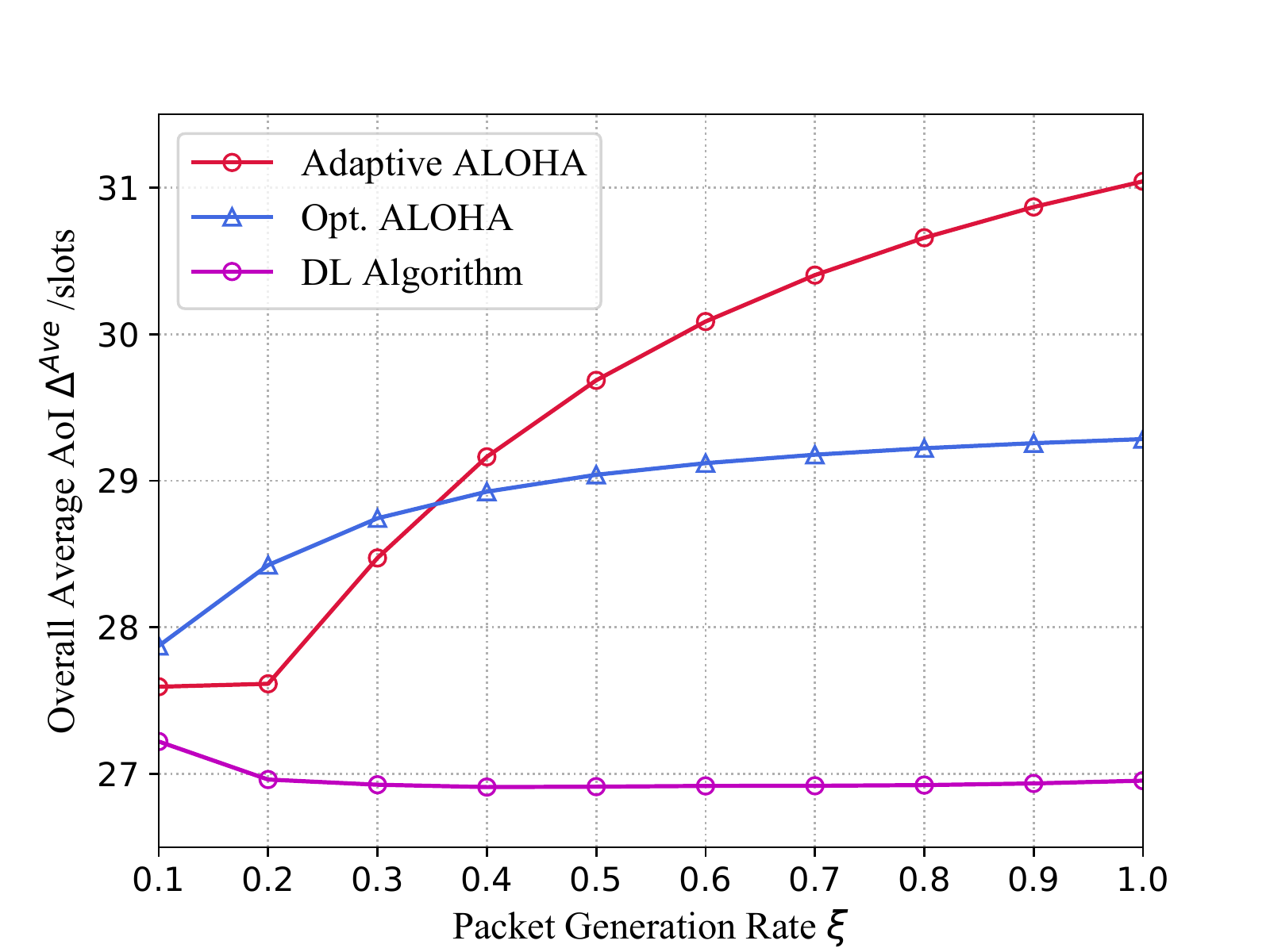}
	\caption{Average $\varDelta^{\mathrm{Ave}}$ performance for $1000$ different D2D layouts as functions of the packet generating rate $\xi$, where we set $\lambda=0$ and $N=400$.}
	\label{fig72}
\end{figure}

In Fig. \ref{fig71} and Fig. \ref{fig72}, we compare the performance of overall average AoI $\varDelta^{\mathrm{Ave}}$ under different method as functions of packet generation rate $\xi$. As depicted, the DL algorithm outperforms Adaptive ALOHA and Opt. ALOHA for diverse $\xi$, and the performance superiority increases with the number of links. Moreover, even with the packet-replacement policy, not the fastest packet generation rate performs the best, which is more evident in Adaptive ALOHA and Opt. ALOHA. This is because faster packet generation rates result in more interference when new packets are always available to transmit over. Infrequent packet generation reduces the buffer non-empty probability, which reduces the level of interference. However, as the packet generating rate further decreases, AoI will be limited by no packet transferability rather than crosslink interference. Besides, it is worth noting that our DL algorithm achieves better performance at high packet generation rates than others.

\subsection{Computation Complexity Analysis}
\begin{figure}[t]
	\centering
	\includegraphics[width=10cm]{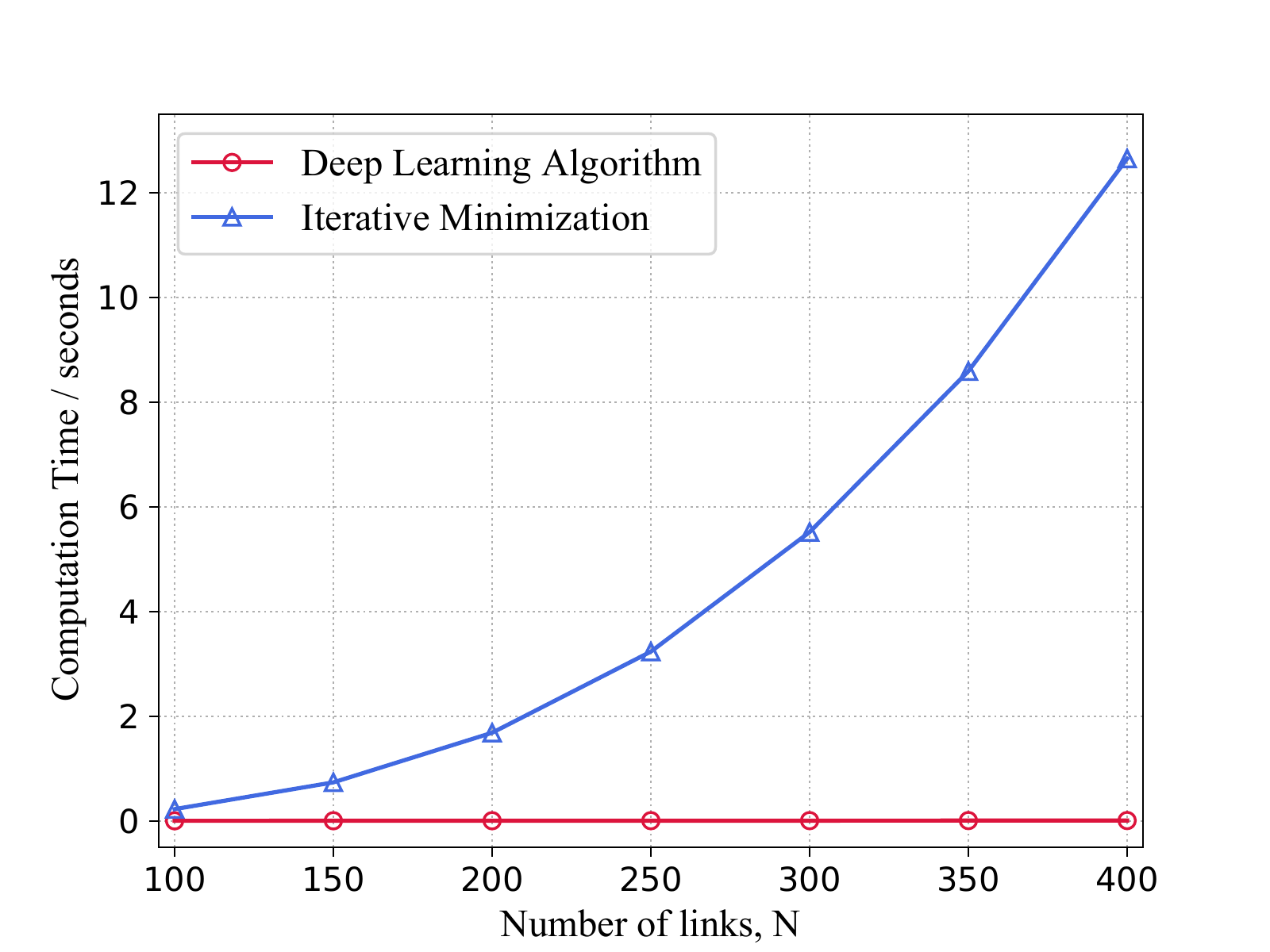}	
	\caption{Computation time consumed to obtain the activation probabilities for one layout in a log scale.}		
	\label{fig8}
\end{figure}
Although numerical results illustrate that the performance of the DL algorithm is slightly inferior to that of the iterative minimization algorithm, the advantage of the DL algorithm mainly lies in its low computational complexity. We analyze the computational complexity of each method we employed.

Iterative minimization algorithm needs to solve $N$ converted one-variable problems in each iteration. The dominating part of the computation of solving each one-variable problem is the matrix operation of $I_{ij}$ and $D_{ij}$. Since the size of the matrix is $N\times N$, the computational complexity of solving this problem is at least $O(N^2)$. Assuming that the number of iterations is a constant independent of $N$, the computational complexity of the iteration minimization algorithm is $O(N^3)$.

After the neural network is well-trained, the computation of the deep learning algorithm to generate scheduling decisions can be divided into the convolution stage and the fully connected stage. Convolution stage needs to complete $M^2 \times(n_1^2+n_2^2+n_3^2)$ times of operation, in which $M$ and $n_i, i=1,2,3$ are the size of GLI grid and the convolution filters respectively. In the fully connected stage, $N$ parallel fully connected networks need to be operated, and $8m_1m_2$ times of computation needs to be completed for each fully connected network, where $8$ represents the length of the input feature vectors and $m_1, m_2$ are the numbers of the neurons for each hidden layer under the utilized neural network configuration. Thus, the total computation is $r\times [M^2\times(n_1^2+n_2^2+n_3^2) + N\times 10m_1m_2]$ where $r$ is the feedback rounds, and the computational complexity of the deep learning algorithm is $O(N)$.

We record the computation time of the two algorithms in generating scheduling decisions, as depicted in Fig. \ref{fig8}. Although the two methods use GPU and CPU resources to obtain the solution respectively and cannot be compared directly, Fig. \ref{fig8} can illustrate the significant difference in computational complexity between the two methods. With the increase of $N$, the deep learning algorithm can achieve a near-optimal performance while possessing the advantage of low computational complexity compared with the iterative approach.

\subsection{Performance of Throughput and AoI Jointly Scheduling}

\begin{figure}[t]
	\centering	
	\includegraphics[width=10cm]{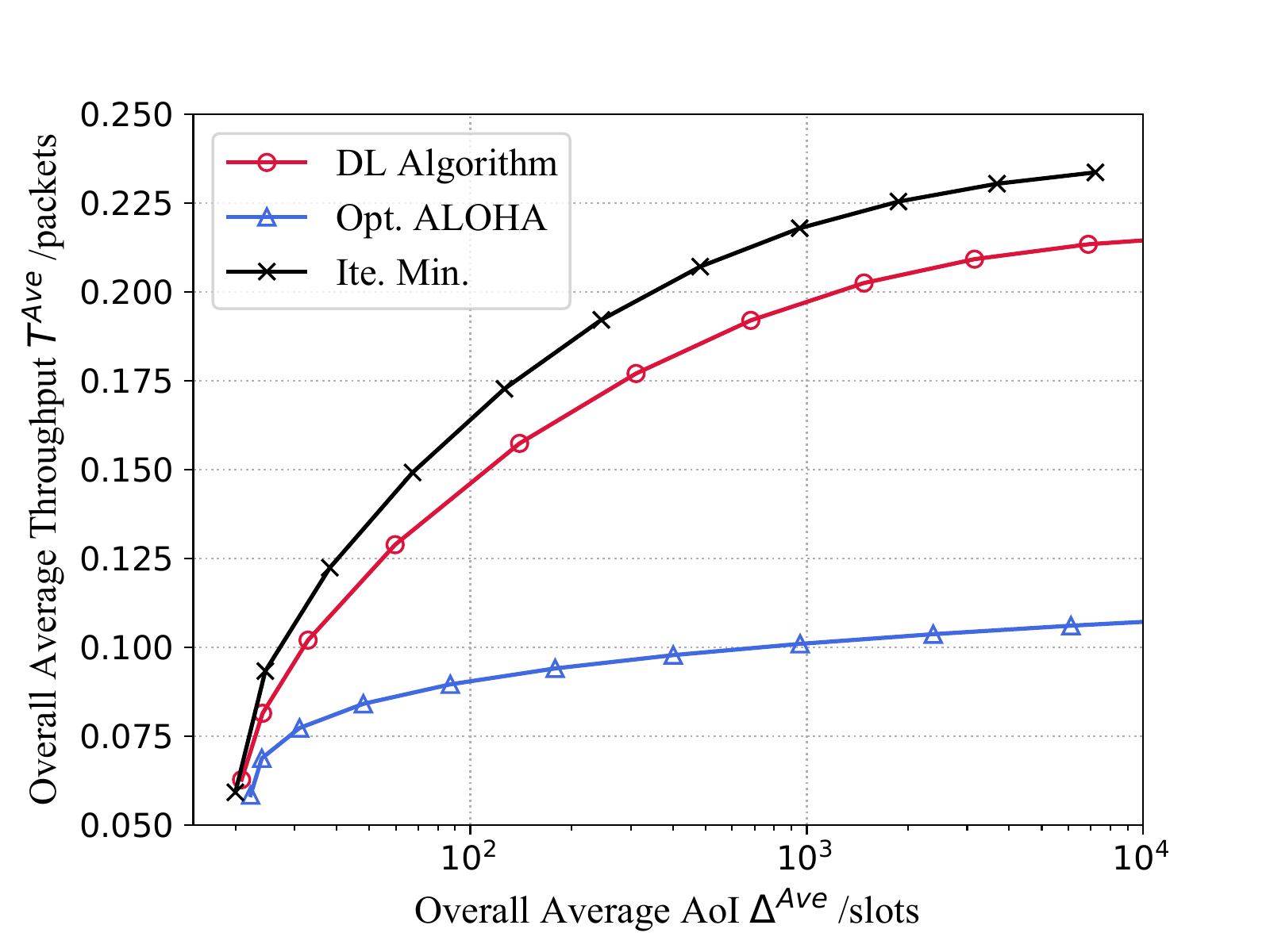}
	\caption{The Pareto front of overall average AoI $\varDelta^{\mathrm{Ave}}$ and throughput $T^{\mathrm{Ave}}$ obtained by different approaches, where we set $N=300$ and $\xi=1$.}
	\label{fig11}
\end{figure}

\begin{figure}[t]
	\centering	
	\includegraphics[width=10cm]{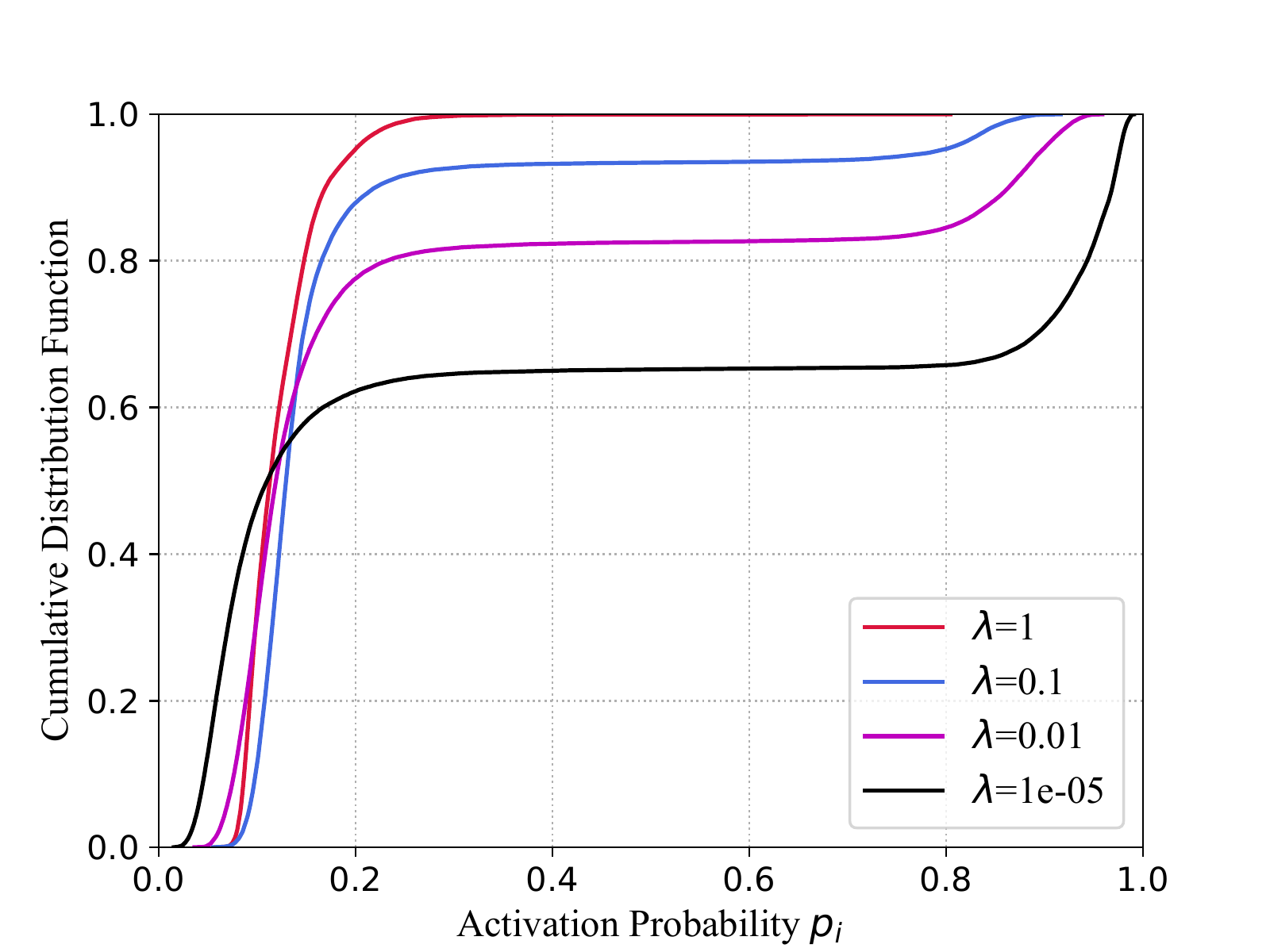}
	\caption{Cumulative distribution of the activation probability $p_i$ for different weight $\lambda$, where we set $N=300$ and $\xi=1$.}	
	\label{fig112}
\end{figure}

\begin{figure}[t]
	\centering	
	\includegraphics[width=10cm]{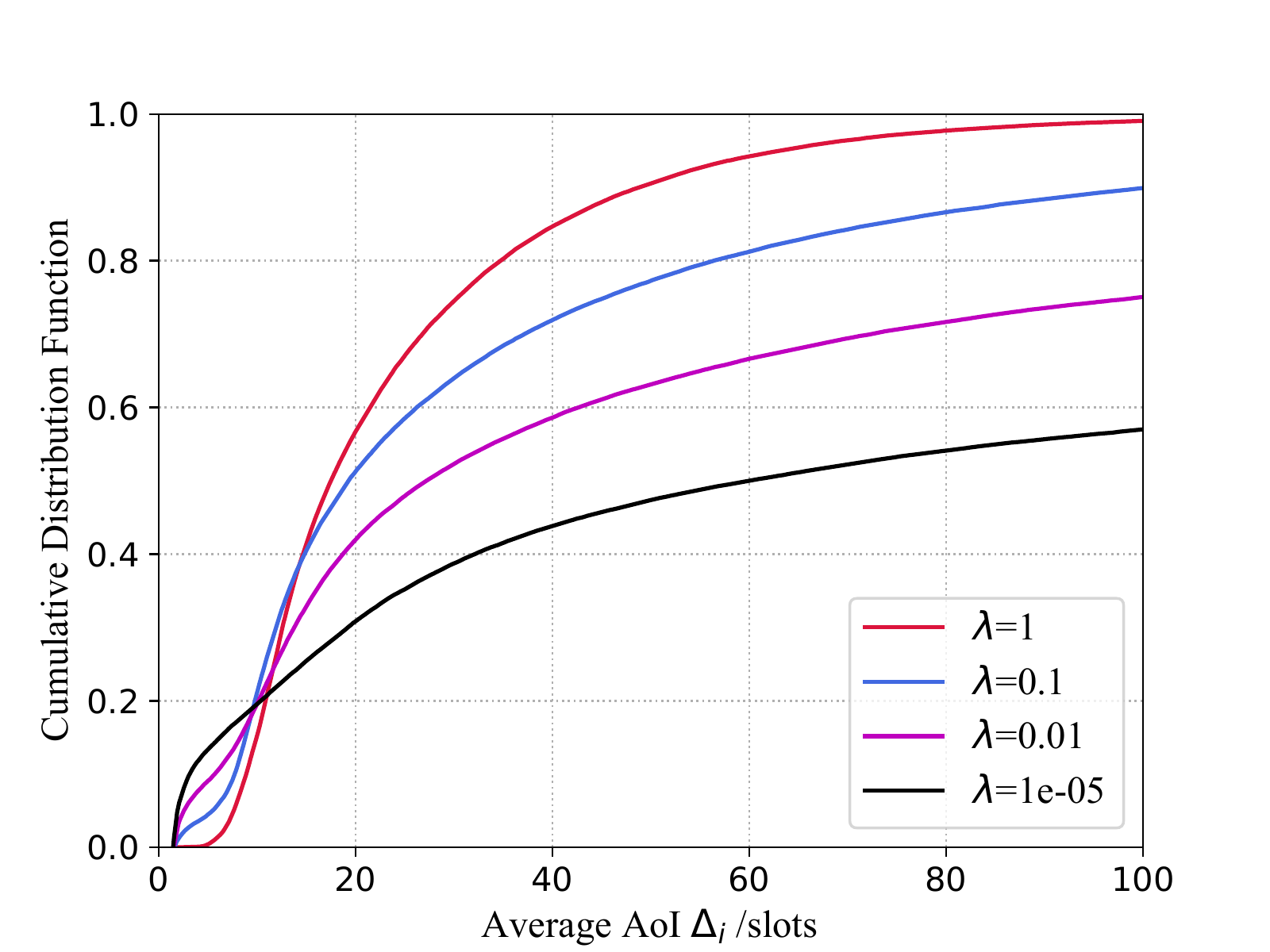}
	\caption{Cumulative distribution of the average AoI $\Delta_i$ for different weight $\lambda$, where we set $N=300$ and $\xi=1$.}
	\label{fig111}
\end{figure}

\begin{figure}[t]
	\centering	
	\includegraphics[width=10cm]{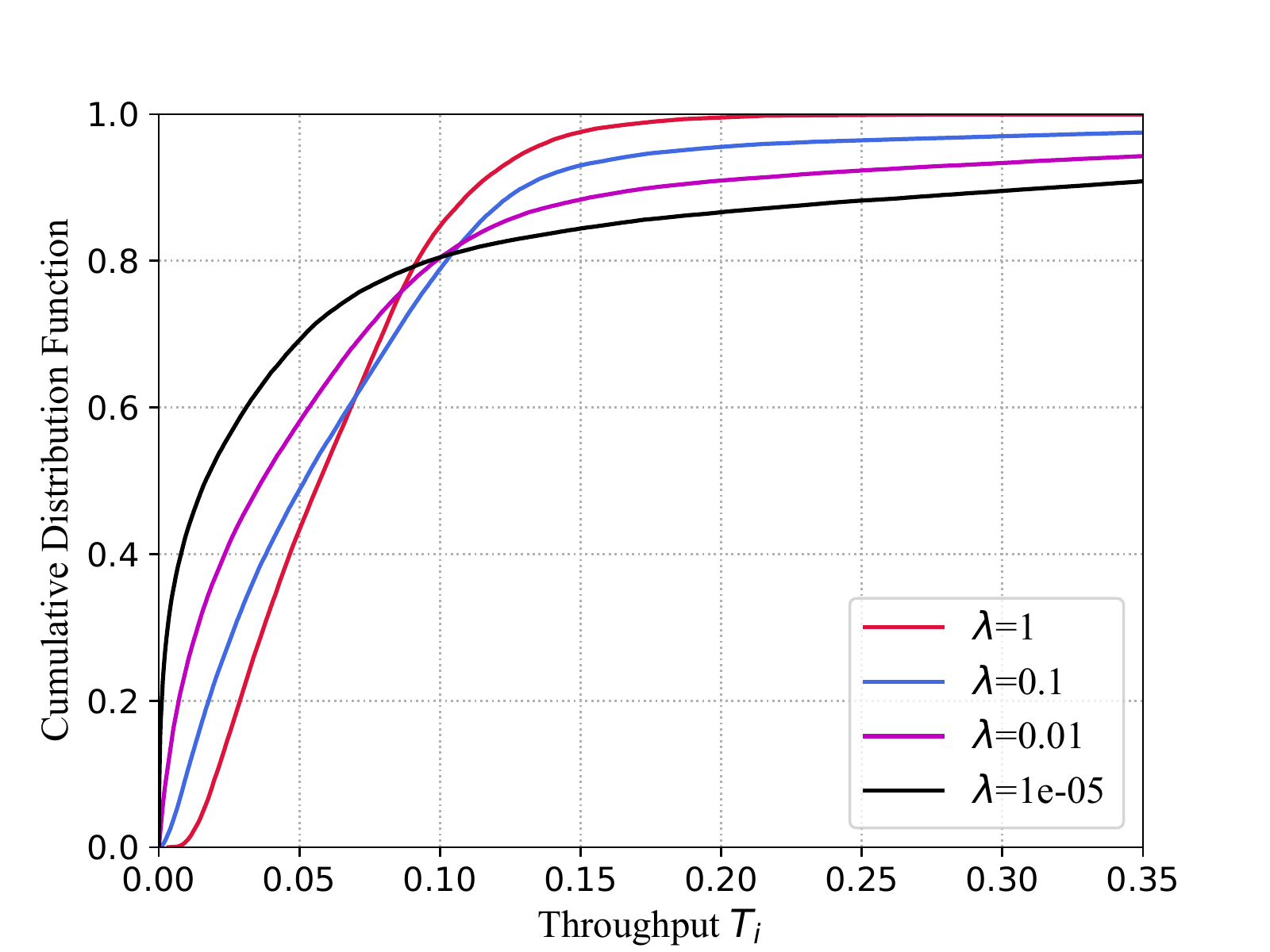}
	\caption{Cumulative distribution of the average throughput $T_i$ for different weight $\lambda$, where we set $N=300$ and $\xi=1$.}
	\label{fig113}
\end{figure}

In the scheduling considering the trade-off between throughput and AoI, we set the number of links to $N=300$ and $\xi=1$. We use Ite. Min. and Opt. ALOHA as a comparison.
Fig.\ref{fig11} depicts the Pareto front of $(\Delta^{Ave},T^{Ave})$ obtained by different methods with the variety of $\lambda$. A network employing great scheduling decisions possesses high throughput and low AoI, thus curves closer to the top left in the figure represent better performance. The DL algorithm has	 approximate performance compared with the iterative minimization algorithm. As depicted in the figure, with the weight of throughput increasing, i.e.,$\lambda$ tends to zero, a little more throughput gain will result in a greater loss of AoI, and AoI tends to infinity when throughput is maximized.
In the case of a single link, there is no trade-off between AOI and throughput, but for large-scale D2D networks, the lowest AOI is obtained in the case of low throughput. This is because the scheduling of throughput maximization will drive the links with great channel conditions to update frequently to achieve better network throughput performance, while the links with poor channel quality will be ignored, resulting in very low update frequency and very high AoI.

Specifically, we illustrate the cumulative ditribution of the activation probability $p_i$, the average AoI $\Delta_i$ and the avearge throughput $T_i$ for links in Fig. \ref{fig112}-\ref{fig113}. We set $N=300$ and $\xi=1$ and $\lambda \in \{1, 0.1, 0.01, 1\text{e-}5\}$.
Fig. \ref{fig113} depicts the allocation of activation probabilities under different AoI weights. When high AoI weights are employed, all links have similar activation probabilities, and the distribution is relatively smooth. Links with low activation probabilities are less than the weights of other conditions. As the AoI weight decreases gradually, the link activation becomes polarized, and the cumulative distribution function has a too sharp increase.
Fig. \ref{fig111} and Fig. \ref{fig113} depict the distribution of AoI and throughput under different AoI weights. Similar to the distribution of activation probability, the higher AoI weight makes the distribution more centralized, and there are fewer links with higher AoI and lower throughput, which implies that the high AoI weight forces the links with poor channel conditions to increase the transmission opportunity. Even if it causes a sacrifice of resources, it pays more attention to the fairness between links than throughput maximizing scheduling.

\section{Conclusion} \label{conclusion}
\textcolor{blue}{In this paper, we considered the link scheduling policy aiming at the joint optimization of AoI and throughput for packets transmitted with an LCFS-PR queue rule for the D2D communication system. We model this system to a spatiotemporal interacting queue and derive the expressions of overall average AoI and throughput based on the mean-field approximation under the randomized stationary policy. A multi-objective optimization problem of minimizing AoI and maximizing throughput simultaneously is formulated under the system model and transformed into a single objective problem by the weighted sum approach. A neural network structure was proposed to directly construct the mapping from GLI to optimal scheduling decisions without estimating CSI, which is trained by unsupervised learning for an implicit objective function.
Numerical simulation obtains the Pareto Front curve between AoI and throughput by utilizing the deep learning algorithm, and reveals that 1) the maximum packet generation rate can not minimize AoI, 2) the throughput-based scheduling makes the link scheduling tend to polarization, while the AoI-based scheduling is fairer.}

\appendix

\section{Appendix}
\subsection{The derivation of $\Delta^{Ave}$ and $T^{Ave}$} \label{app1}
The average AoI for a single transmission link employing LCFS G/G/1 queue with preemptive
service has been derived in \cite{tripathi2019age} as follows
\begin{align}
	\Delta_i = \frac{1}{2} \frac{\mathbb{E}[X_i^2]}{\mathbb{E}[X_i]} + \frac{\mathbb{E}[\min(X_i,S_i)]}{\mathbb{P}[S_i \leq X_i]} - \frac{1}{2}.\label{aoi}
\end{align}

Owing to $X_i$ and $S_i$ are both geometric distributions, we can easily obtain the following derivation \cite{yang2021spatiotemporal}:
\begin{align}
	&\mathbb{E}[X_i] = \frac{1}{\xi}, \ \ \mathbb{E}[X_i^2] = \frac{2-\xi}{\xi^2} \label{11} \\
	&\mathbb{E}[\min(X_i,S_i)] = \frac{1}{\xi + (1 - \xi)p_i\mu_i} \label{22} \\
	&\mathbb{P}[S_i \leq X_i] = 1 - \mathbb{E}[(1-p_i\mu_i)^{X_i}] = \frac{p_i\mu_i}{\xi + (1 - \xi)p_i\mu_i}.\label{33}
\end{align}

Substituting \eqref{11}-\eqref{33} into \eqref{aoi}, we can derive the average AoI $\Delta_i$ as follows:
\begin{align}
	\Delta_i = \frac{1}{\xi} + \frac{1}{p_i\mu_i}-1 \label{a_i}
\end{align}

The throughput of each link $T_i$ can also be derived under the mean-field assumption:
\begin{align}
	T_i &= \lim_{K \to\infty}\mathbb{E}\left[b_i(k)\right] = \lim_{K \to\infty}\mathbb{E}\left[a_j(k)n_j(k)\mu_j(k)\right] \nonumber\\
	&= p_i\nu_i\mu_i = \frac{\xi p_i \mu_i}{\xi + (1 - \xi)p_i\mu_i}. \label{thr_i}
\end{align}

The results in \eqref{aoi_all} and \eqref{thr_all} follows from substituting \eqref{a_i} and \eqref{thr_i} into \eqref{e6} and \eqref{e66}.

\subsection{The derivation of $\frac{\partial\mu_j}{\partial p_i}$} \label{appb}
Fistly, substituting \eqref{nu} into \eqref{mupi}, and calculating logarithmic functions on both sides of the equation
\begin{align}
	\ln\mu_i &= \ln\rho_i + \sum_{j \neq i}\ln\left(1 - \frac{p_j}{1+D_{ji}}\frac{\xi}{\xi+(1-\xi)p_i\mu_i}\right) \\
	&= \ln\rho_i + \sum_{j \neq i}\ln\left(1 - \frac{p_j B_{ji}}{1+q p_i \mu_i}\right), \label{eq2:lnmu}
\end{align}
where $B_{ij}=\frac{1}{1 + D_{ij}}$ and $q = \frac{1-\xi}{\xi}$. Then, calculating the partial derivative of \eqref{eq2:lnmu} to $p_m$. $m$ is used to distinguish the two subscripts $i$ and $j$. The following equations can be derived
\begin{align}
	\frac{1}{\mu_i}\frac{\partial\mu_i}{\partial p_m} &= -\sum_{j \neq i} \frac{\frac{\partial p_j}{\partial p_m}B_{ji}-\frac{\partial\mu_j}{\partial p_m}qB_{ji}p_j^{2}}{(1+q p_j \mu_j)(1+qp_j\mu_j-B_{ji}p_i)} \\
	&= - \frac{B_{mi}}{(1+q p_m \mu_m)(1+qp_m\mu_i-B_{mi}p_m)} + \sum_{j \neq i} \frac{\frac{\partial\mu_j}{\partial p_m}qB_{ji}p_j^{2}}{(1+q p_j \mu_j)(1+qp_j\mu_j-B_{ji}p_j)} . \label{eq2:partial}
\end{align}

Computing\eqref{eq2:partial} for each $i$ and sort out these $N$ equations, we can obtain
\begin{align}
	P\bm{\mu}_{p_m}^T &= \bm{q}_m \label{eq2:mat}  \\
	\bm{\mu}_{p_m}&=[\frac{\partial\mu_1}{\partial p_m},\frac{\partial\mu_2}{\partial p_m},\dots,\frac{\partial\mu_N}{\partial p_m}] \\
	P_{ji} &= \frac{qB_{ij}p_i^2}{(1+qp_i\mu_i)(1+qp_i\mu_i-B_{ij}p_i)}, j\neq i,\ \  P_{(j,j)} = -\frac{1}{\mu_j}, \label{eq2:pji} \\
	q_{m,i} &= \frac{B_{mi}}{(1+qp_m\mu_m)(1+qp_m\mu_i-B_{mi}p_m)} ,
\end{align}
where $(\cdot)^T$ represents matrix transpose and $\bm{\mu}_{p_m}$ represents the partial derivative of $\{\mu_i\}_{i=1}^N$ to $p_m$. $P_{ij}$ is the element of $P$ at index $ij$ and $q_{m,i}$ is the element of $\bm{q}_m$ at index $i$.

Computing \eqref{eq2:mat} for each $m$ and sort out these $N$ vector equations, we can obtain
\begin{align}
	PM &= Q, \label{eq2:matr}  \\
	M &= [\bm{\mu}_{p_1},\bm{\mu}_{p_2}, \dots ,\bm{\mu}_{p_N}],  \\
	Q &= [\bm{q}_1,\bm{q}_2, \dots ,\bm{q}_N], \\
	Q_{ji} &= \frac{B_{ij}}{(1+qp_i\mu_i)(1+qp_i\mu_i-B_{ij}p_i)}. \label{eq2:qji}
\end{align}

Finally, $\frac{\partial\mu_j}{\partial p_i}$ can be obtained by matrix multiplication as follows
\begin{align}
	\frac{\partial\mu_j}{\partial p_i} = M_{ji} = (P^{-1}Q)_{ji} ,
\end{align}
where $(\cdot)^{-1}$ represents inverse matrix and \eqref{eq2:pji} and \eqref{eq2:qji} is \eqref{Mji} and \eqref{Nji}

\bibliographystyle{IEEEtran}
\bibliography{AoI}

\begin{thebibliography}{10}
\providecommand{\url}[1]{#1}
\csname url@samestyle\endcsname
\providecommand{\newblock}{\relax}
\providecommand{\bibinfo}[2]{#2}
\providecommand{\BIBentrySTDinterwordspacing}{\spaceskip=0pt\relax}
\providecommand{\BIBentryALTinterwordstretchfactor}{4}
\providecommand{\BIBentryALTinterwordspacing}{\spaceskip=\fontdimen2\font plus
\BIBentryALTinterwordstretchfactor\fontdimen3\font minus
  \fontdimen4\font\relax}
\providecommand{\BIBforeignlanguage}[2]{{%
\expandafter\ifx\csname l@#1\endcsname\relax
\typeout{** WARNING: IEEEtran.bst: No hyphenation pattern has been}%
\typeout{** loaded for the language `#1'. Using the pattern for}%
\typeout{** the default language instead.}%
\else
\language=\csname l@#1\endcsname
\fi
#2}}
\providecommand{\BIBdecl}{\relax}
\BIBdecl

\bibitem{2021mobile}
\BIBentryALTinterwordspacing
G.~Intelligence, ``The mobile economy 2021,'' \emph{London: GSM Association},
  2021. [Online]. Available: \url{https://www.gsma.com/mobileeconomy/}
\BIBentrySTDinterwordspacing

\bibitem{FPLinQ}
K.~{Shen} and W.~{Yu}, ``Fplinq: A cooperative spectrum sharing strategy for
  device-to-device communications,'' in \emph{Proc. IEEE International
  Symposium on Information Theory (ISIT)}, Jun. 2017, pp. 2323--2327.

\bibitem{CachingPolicy}
M.~{Lee} and A.~F. {Molisch}, ``Caching policy and cooperation distance design
  for base station-assisted wireless d2d caching networks: Throughput and
  energy efficiency optimization and tradeoff,'' \emph{IEEE Transactions on
  Wireless Communications}, vol.~17, no.~11, pp. 7500--7514, Nov. 2018.

\bibitem{Heuristic}
J.~{Gu}, S.~J. {Bae}, S.~F. {Hasan}, and M.~Y. {Chung}, ``Heuristic algorithm
  for proportional fair scheduling in d2d-cellular systems,'' \emph{IEEE
  Transactions on Wireless Communications}, vol.~15, no.~1, pp. 769--780, Jan.
  2016.

\bibitem{8688635}
G.~{Chisci}, H.~{Elsawy}, A.~{Conti}, M.~{Alouini}, and M.~Z. {Win},
  ``Uncoordinated massive wireless networks: Spatiotemporal models and
  multiaccess strategies,'' \emph{IEEE/ACM Transactions on Networking},
  vol.~27, no.~3, pp. 918--931, Jun. 2019.

\bibitem{Real-time}
S.~{Kaul}, R.~{Yates}, and M.~{Gruteser}, ``Real-time status: How often should
  one update?'' in \emph{Proc. IEEE INFOCOM}, Mar. 2012, pp. 2731--2735.

\bibitem{OntheAge}
M.~{Costa}, M.~{Codreanu}, and A.~{Ephremides}, ``On the age of information in
  status update systems with packet management,'' \emph{IEEE Transactions on
  Information Theory}, vol.~62, no.~4, pp. 1897--1910, Apr. 2016.

\bibitem{Update}
Y.~{Sun}, E.~{Uysal-Biyikoglu}, R.~D. {Yates}, C.~E. {Koksal}, and N.~B.
  {Shroff}, ``Update or wait: How to keep your data fresh,'' \emph{IEEE
  Transactions on Information Theory}, vol.~63, no.~11, pp. 7492--7508, Nov.
  2017.

\bibitem{8323423}
C.~{Kam}, S.~{Kompella}, G.~D. {Nguyen}, J.~E. {Wieselthier}, and
  A.~{Ephremides}, ``On the age of information with packet deadlines,''
  \emph{IEEE Transactions on Information Theory}, vol.~64, no.~9, pp.
  6419--6428, Sept. 2018.

\bibitem{8445919}
B.~{Wang}, S.~{Feng}, and J.~{Yang}, ``To skip or to switch? minimizing age of
  information under link capacity constraint,'' in \emph{Proc. IEEE 19th
  International Workshop on Signal Processing Advances in Wireless
  Communications (SPAWC)}, Jun. 2018, pp. 1--5.

\bibitem{8469047}
R.~D. {Yates} and S.~K. {Kaul}, ``The age of information: Real-time status
  updating by multiple sources,'' \emph{IEEE Transactions on Information
  Theory}, vol.~65, no.~3, pp. 1807--1827, Mar. 2019.

\bibitem{8437907}
R.~D. {Yates}, ``Status updates through networks of parallel servers,'' in
  \emph{Proc. IEEE International Symposium on Information Theory (ISIT)}, Jun.
  2018, pp. 2281--2285.

\bibitem{8006541}
R.~D. {Yates}, E.~{Najm}, E.~{Soljanin}, and J.~{Zhong}, ``Timely updates over
  an erasure channel,'' in \emph{Proc. IEEE International Symposium on
  Information Theory (ISIT)}, Jun. 2017.

\bibitem{10.1145/3397166.3409125}
A.~M. Bedewy, Y.~Sun, R.~Singh, and N.~B. Shroff, ``Optimizing information
  freshness using low-power status updates via sleep-wake scheduling,'' in
  \emph{Proc. of ACM MobiHoc}, Oct. 2020, pp. 51--60.

\bibitem{Scheduling}
I.~{Kadota}, A.~{Sinha}, and E.~{Modiano}, ``Scheduling algorithms for
  optimizing age of information in wireless networks with throughput
  constraints,'' \emph{IEEE/ACM Transactions on Networking}, vol.~27, no.~4,
  pp. 1359--1372, Aug. 2019.

\bibitem{mankar}
\BIBentryALTinterwordspacing
P.~D. Mankar, M.~A. Abd-Elmagid, and H.~S. Dhillon, ``Spatial distribution of
  the mean peak age of information in wireless networks,'' 2020. [Online].
  Available: \url{https://arxiv.org/abs/2006.00290}
\BIBentrySTDinterwordspacing

\bibitem{OptimizingH}
H.~H. {Yang}, A.~{Arafa}, T.~Q.~S. {Quek}, and V.~{Poor}, ``Optimizing
  information freshness in wireless networks: A stochastic geometry approach,''
  \emph{IEEE Transactions on Mobile Computing}, pp. 1--1, 2020.

\bibitem{9042825}
M.~{Emara}, H.~{Elsawy}, and G.~{Bauch}, ``A spatiotemporal model for peak aoi
  in uplink iot networks: Time versus event-triggered traffic,'' \emph{IEEE
  Internet of Things Journal}, vol.~7, no.~8, pp. 6762--6777, Aug. 2020.

\bibitem{7492912}
Q.~{He}, D.~{Yuan}, and A.~{Ephremides}, ``Optimizing freshness of information:
  On minimum age link scheduling in wireless systems,'' in \emph{Proc. 14th
  International Symposium on Modeling and Optimization in Mobile, Ad Hoc, and
  Wireless Networks (WiOpt)}, May 2016, pp. 1--8.

\bibitem{HORNIK1989359}
K.~Hornik, M.~Stinchcombe, and H.~White, ``Multilayer feedforward networks are
  universal approximators,'' \emph{Neural Networks}, vol.~2, no.~5, pp.
  359--366, Mar. 1989.

\bibitem{8664604}
W.~{Cui}, K.~{Shen}, and W.~{Yu}, ``Spatial deep learning for wireless
  scheduling,'' \emph{IEEE Journal on Selected Areas in Communications},
  vol.~37, no.~6, pp. 1248--1261, June. 2019.

\bibitem{9376717}
E.~T. {Ceran}, D.~{Gündüz}, and A.~{György}, ``A reinforcement learning
  approach to age of information in multi-user networks with harq,'' \emph{IEEE
  Journal on Selected Areas in Communications}, pp. 1--1, 2021.

\bibitem{9097584}
B.~{Yin}, S.~{Zhang}, and Y.~{Cheng}, ``Application-oriented scheduling for
  optimizing the age of correlated information: A
  deep-reinforcement-learning-based approach,'' \emph{IEEE Internet of Things
  Journal}, vol.~7, no.~9, pp. 8748--8759, Sept. 2020.

\bibitem{wang2021distributed}
S.~Wang, M.~Chen, Z.~Yang, C.~Yin, W.~Saad, S.~Cui, and H.~V. Poor,
  ``Distributed reinforcement learning for age of information minimization in
  real-time iot systems,'' 2021.

\bibitem{yang2021spatiotemporal}
\BIBentryALTinterwordspacing
H.~H. Yang, A.~Arafa, T.~Q.~S. Quek, and H.~V. Poor, ``Spatiotemporal analysis
  for age of information in random access networks under last-come first-serve
  with replacement protocol,'' 2021. [Online]. Available:
  \url{https://arxiv.org/abs/2109.08825}
\BIBentrySTDinterwordspacing

\bibitem{5061954}
I.~. {Hou}, V.~{Borkar}, and P.~R. {Kumar}, ``A theory of qos for wireless,''
  in \emph{Proc. IEEE INFOCOM}, Apr. 2009, pp. 486--494.

\bibitem{8340813}
F.~Jameel, Z.~Hamid, F.~Jabeen, S.~Zeadally, and M.~A. Javed, ``A survey of
  device-to-device communications: Research issues and challenges,'' \emph{IEEE
  Communications Surveys \& Tutorials}, vol.~20, no.~3, pp. 2133--2168, Apr.
  2018.

\bibitem{8943134}
R.~{Talak}, S.~{Karaman}, and E.~{Modiano}, ``Optimizing information freshness
  in wireless networks under general interference constraints,'' \emph{IEEE/ACM
  Transactions on Networking}, vol.~28, no.~1, pp. 15--28, Feb. 2020.

\bibitem{MOEA}
Q.~{Zhang} and H.~{Li}, ``Moea/d: A multiobjective evolutionary algorithm based
  on decomposition,'' \emph{IEEE Transactions on Evolutionary Computation},
  vol.~11, no.~6, pp. 712--731, Dec. 2007.

\bibitem{15561801}
A.~P. Charles~Bordenave, David R.~McDonald, ``A particle system in interaction
  with a rapidly varying environment: Mean field limits and applications,''
  \emph{Networks \& Heterogeneous Media}, vol.~5, no.~1, pp. 31--62, Mar. 2010.

\bibitem{ITU}
\emph{Recommendation ITU-R P.1411-10}, Int. Telecommun. Union, Geneva,
  Switzerland, 2019.

\bibitem{8186962}
F.~Baccelli and B.~Blaszczyszyn, 2010.

\bibitem{tripathi2019age}
\BIBentryALTinterwordspacing
V.~Tripathi, R.~Talak, and E.~Modiano, ``Age of information for discrete time
  queues,'' 2019. [Online]. Available: \url{https://arxiv.org/abs/1901.10463}
\BIBentrySTDinterwordspacing

\end{thebibliography}

\end{document}